\documentclass[12pt]{article}
\usepackage{amsfonts}
\usepackage{graphicx}
\font\twozero=cmr10 at 20pt
\font\oneeight=cmr10 at 18pt
\font\onesix=cmr10 at 16pt
\font\onefour=cmr10 at 14pt

\newcommand{\vT}{\vphantom{\mbox{\twozero I}}}
\newcommand{\vTm}{\vphantom{\mbox{\oneeight I}}}
\newcommand{\vTmm}{\vphantom{\mbox{\onesix I}}}
\newcommand{\vTs}{\vphantom{\mbox{\onefour I}}}

\begin{document}

\baselineskip=20pt

\newfont{\elevenmib}{cmmib10 scaled\magstep1}
\newcommand{\preprint}{
   \begin{flushleft}
     \elevenmib Yukawa\, Institute\, Kyoto\\
   \end{flushleft}\vspace{-1.3cm}
   \begin{flushright}\normalsize  \sf
     YITP-03-51\\
     {\tt hep-th/0308052} \\ August 2003
   \end{flushright}}
\newcommand{\Title}[1]{{\baselineskip=26pt
   \begin{center} \Large \bf #1 \\ \ \\ \end{center}}}
\newcommand{\Author}{\begin{center}
   \large  I.~Loris${}^{a,b}$ and R.~Sasaki${}^a$ \end{center}}
\newcommand{\Address}{\begin{center}
     ${}^a$ Yukawa Institute for Theoretical Physics,\\
     Kyoto University, Kyoto 606-8502, Japan\\
${}^b$ Dienst Theoretische Natuurkunde,
     Vrije Universiteit Brussel,   \\ 
Pleinlaan 2, B-1050 Brussels,  Belgium
   \end{center}}
\newcommand{\Accepted}[1]{\begin{center}
   {\large \sf #1}\\ \vspace{1mm}{\small \sf Accepted for Publication}
   \end{center}}

\preprint
\thispagestyle{empty}
\bigskip\bigskip\bigskip

\Title{ Quantum \& Classical  Eigenfunctions in \\
Calogero \& Sutherland  Systems }
\Author

\Address
\vspace{1cm}

\begin{abstract}
An interesting observation was reported by Corrigan-Sasaki that
all the frequencies of small oscillations around equilibrium are ``{\em
quantised\/}" for Calogero and Sutherland (C-S) systems, 
typical integrable multi-particle dynamics.
We present an analytic proof by applying  recent results of Loris-Sasaki.
Explicit forms of `classical' and quantum 
eigenfunctions are presented for C-S systems
based on any root systems. 
\end{abstract}

\section{Introduction}
\label{intro}
\setcounter{equation}{0}
In a recent paper \cite{ls1} simple theorems pertaining to the
correspondence between  quantum and classical dynamics are proved for the
general multi-particle quantum mechanical systems with discrete eigenvalues.
The theorems relate quantum mechanical eigenvalues and eigenfunctions to
the properties of the classical mechanical system at equilibrium.
Corresponding to each quantum eigenfunction, a `{\em classical eigenfunction\/}'
is defined whose eigenvalue is given by the `main part', 
that is the order $\hbar$ part,
of the quantum eigenvalue.
For the `{\em elementary excitations\/}' these classical and quantum eigenvalues
are nothing but the eigenfrequencies of the normal modes of the small
oscillations at equilibrium.

We apply these theorems to the Calogero and Sutherland \cite{Cal}
systems, typical integrable multi-particle dynamics with long range interactions 
based on root  systems \cite{OP1}.
The theorems provide an analytic proof for the interesting observations
made by Corrigan-Sasaki \cite{cs,perepoin,perenew} concerning the classical and quantum
integrability in Calogero and Sutherland systems.
Explicit forms of the classical and quantum eigenfunctions for the
elementary excitations are presented for  the Calogero and Sutherland systems
based on any root systems.
These exemplify another aspect of the close relationship
between the classical and quantum integrability in Calogero and Sutherland
systems.

This paper is organised as follows.
In section two,  basic formulation of multi-particle quantum mechanics
in terms of the {\em prepotential\/} \cite{bms,kps} is briefly reviewed.
After the reformulation of the quantum mechanical wavefunctions at
equilibrium, the main theorem of the Loris-Sasaki paper \cite{ls1}
is recapitulated.
In section three the basic concepts of the Calogero and Sutherland systems
are summarised.
Section four and five are the main part of this paper,
presenting the classical and quantum eigenfunctions
of the Calogero systems (section four) and Sutherland systems (section five).
The final section is for summary and comments.

\section{Basic Quantum Mechanics}
\label{qmechanics}
\setcounter{equation}{0}

Let us start with a basic formulation of 
multi-particle quantum mechanical
system in terms of a {\em prepotential} \cite{bms,kps} and 
later we will discuss its relationship with the corresponding
classical ($\hbar\to0$) dynamics. 
The dynamical variables 
are the coordinates
\(\{q_{j}|\,j=1,\ldots,r\}\) and their canonically conjugate momenta
\(\{p_{j}|\,j=1,\ldots,r\}\), subject to the Heisenberg commutation relations 
or the Poisson
bracket relations:
\begin{eqnarray*}
 [q_{j},p_{k}]=i\hbar\delta_{j\,k},\qquad [q_{j},q_{k}]=
   [p_{j},p_{k}]=0,\\
 \{q_{j},p_{k}\}=\delta_{j\,k},\qquad \{q_{j},q_{k}\}=
   \{p_{j},p_{k}\}=0.
\end{eqnarray*}
We will adopt the standard vector notation in  \(\mathbf{R}^{r}\):
\begin{equation}
   q=(q_{1},\ldots,q_{r}),\quad p=(p_{1},\ldots,p_{r}),
\quad q^2\equiv\sum_{j=1}^rq_j^2, 
\quad p^2\equiv\sum_{j=1}^rp_j^2, \ldots,
\label{qpdefs}
\end{equation}
in which $r$ is the number of particles.
In quantum theory, the momentum operator \(p_j\) acts as
a differential operator:
\[
   p_j=-i\hbar{\partial\over{\partial q_j}}, \quad j=1,\ldots,r.
\]

\bigskip
Throughout this paper we discuss the standard Hamiltonian system
\begin{equation}
H={1\over2}p^2+V(q),
\end{equation}
in which  we have assumed for simplicity that 
all the particles have the same mass,
which is rescaled to unity.
Let us start with  mild assumptions that the system has a 
unique and {\em square integrable\/} ground state $\psi_0$:
\begin{equation}
H\psi_0=0,\quad \int|\psi_0|^2\,d^r\!q<\infty,
\end{equation}
and that it has  a finite (or an infinite) number of
{\em
discrete\/} eigenvalues:
\begin{equation}
H\psi_n=E_n\psi_n,\quad
E_n={\mathcal E}_n\hbar +\mathcal{O}(\hbar^2).
\label{eigexp}
\end{equation}
Here we adopt the convention that the ground state energy is vanishing, by
adjusting the constant part of the potential $V$, see below.

Since the above time-independent Schr\"odinger equation is real
for a self-adjoint Hamiltonian and that the ground state has no {\em node\/} 
we express the ground state eigenfunction as
\begin{equation}
\psi_0(q)=e^{{1\over\hbar}W(q)},
\label{grstate}
\end{equation}
in which a real function $W=W(q)$ is called a {\em prepotential\/}
\cite{bms,kps}. By simple differentiation of (\ref{grstate}), we obtain
\begin{equation}
p_j\psi_0=-i{\partial W\over{\partial q_j}}\psi_0,\quad
p^2\psi_0=-\sum_{j=1}^r\left[\left({\partial W\over{\partial q_j}}\right)^2
+\hbar{\partial^2 W\over{\partial q_j^2}}\right]\psi_0,
\end{equation}
which results in 
\begin{equation}
\left\{{1\over2}p^2+{1\over2}\sum_{j=1}^r\left[\left({\partial W\over{\partial
q_j}}\right)^2 +\hbar{\partial^2 W\over{\partial q_j^2}}\right]\right\}\psi_0=0.
\end{equation}
In other words, we can express the potential (plus the ground state energy)
in terms of the prepotential \cite{cs,bms,kps}%
\footnote{
Similar formulas can be found within the context 
of supersymmetric quantum mechanics
\cite{khare}. Here we stress that supersymmetry is not necessary.}
\begin{equation}
V(q)={1\over2}\sum_{j=1}^r\left[\left({\partial W\over{\partial q_j}}\right)^2
+\hbar{\partial^2 W\over{\partial q_j^2}}\right].
\label{potform}
\end{equation}
By removing the obvious $\hbar$-dependent terms,  let us define a {\em
classical\/} potential
$V_C(q)$:
\begin{equation}
V_C(q)={1\over2}\sum_{j=1}^r\left({\partial W\over{\partial q_j}}
\right)^2.
\label{clpot}
\end{equation}
Conversely, (\ref{potform}) is a Riccati equation  determining the
prepotential $W$ for a given potential $V$ (or $V_C$).
Needless to say, it does not matter 
if the prepotential can be expressed in terms of
elementary functions or not.


\subsection{Equilibrium position and frequencies of small oscillations}

Now let us consider the equilibrium point of the  classical potential
$V_C$ (\ref{clpot}). The classical Hamiltonian $H_C={p^2/2}+V_C$ has a stationary
solution
at the classical equilibrium point, $p=0$,\ $q=\bar{q}$.
There could be, in general,  many stationary points 
of the classical potential $V_C$,
among which we will focus on the `{\em maximum}' point  $\bar{q}$
of the ground state
wavefunction $\psi_0$ \cite{cs}:
\begin{equation}
\left.{\partial W\over{\partial
q_j}}\right|_{\bar{q}}=0,\qquad \Longrightarrow
\left.{\partial V_C\over{\partial
q_j}}\right|_{\bar{q}}=\sum_{k=1}^r\left.
{\partial^2 W\over{\partial q_j\partial q_k}}\right|_{\bar{q}}\left.{\partial
W\over{\partial q_k}}\right|_{\bar{q}}=0,\quad j=1,\ldots, r.
\label{Vmin}
\end{equation}
By expanding the classical potential $V_C$ around $\bar{q}$
(\ref{Vmin}), we obtain
\begin{eqnarray}
V_C(q)&=&{1\over2}\sum_{j,\,k=1}^r
\left.{\partial^2 V_C\over{\partial q_j\partial
q_k}}\right|_{\bar{q}}(q-\bar{q})_j(q-\bar{q})_k +\mathcal{O}((q-\bar{q})^3)
\nonumber\\
&=&{1\over2}\sum_{j,\,k,\,l=1}^r\left.{\partial^2 W\over
{\partial q_j\partial q_l}}\right|_{\bar{q}}
\left.{\partial^2 W\over{\partial q_l\partial
q_k}}\right|_{\bar{q}}(q-\bar{q})_j(q-\bar{q})_k +\mathcal{O}((q-\bar{q})^3),
\end{eqnarray}
since $V_C(\bar{q})=0$, (\ref{clpot}).
Thus the eigen (angular) frequencies ((frequency)${}^2$) of small oscillations
near the classical equilibrium are given as the eigenvalues of the Hessian matrix
$\widetilde{W}$ ($\widetilde{V}_C$):
\begin{equation}
\widetilde{W}=\mbox{Matrix}\left[\left.\ 
{\partial^2 W\over{\partial q_j\partial
q_k}}\right|_{\bar{q}}\right],\quad 
\widetilde{V}_C=\mbox{Matrix}\left[\left.\ 
{\partial^2 V_C\over{\partial q_j\partial
q_k}}\right|_{\bar{q}}\right]=\widetilde{W}^2.
\label{VWmat}
\end{equation}


\subsection{Classical Limit of Quantum Eigenfunctions}
\label{atsuteigfun}

Let us express the discrete eigenfunctions in  product forms
\begin{equation}
\psi_n(q)=\phi_n(q)\psi_0(q),\quad n=0,1,\ldots,
\qquad \phi_0\equiv1,
\end{equation}
in which $\phi_n$ obeys a simplified equation with the similarity transformed
Hamiltonian $\hat{H}$ \cite{bms,kps}:
\begin{eqnarray}
\hat{H}\phi_n&=&E_n\phi_n,\label{sthameq}\\
\hat{H}=e^{-{1\over\hbar}W}H e^{{1\over\hbar}W}
&=&-{\hbar^2\over2}\triangle +\hbar\hat{A},
\label{htilform}
\end{eqnarray}
in which $\triangle$ is the Laplacian and a linear differential
operator $\hat{A}$ is
defined for any smooth function
$\varphi(q)$ as:
\begin{equation}
(\hat{A}\varphi)(q)\equiv -\sum_{j=1}^r{\partial W(q)\over{\partial
q_j}}{\partial
\varphi(q)\over{\partial q_j}},\qquad 
\triangle=\sum_{j=1}^r{\partial^2\over{\partial q_j^2}}.
\label{cleigen}
\end{equation}
Here we adjust the normalisation of  the eigenfunctions
$\{\phi_n\}$ so that the corresponding ``{\em classical\/}"
eigenfunctions $\{\varphi_n\}$ are finite (non-vanishing) 
in the limit $\hbar\to0$:
\begin{equation}
\lim_{\hbar\to0}\phi_n(q)=\varphi_n(q),\quad n=1,2,\ldots, .
\label{clasdef}
\end{equation}
By taking the classical limit ($\hbar\to0$) of (\ref{sthameq}) and considering
(\ref{eigexp}), (\ref{htilform}), we arrive at an `{\em eigenvalue equation\/}' 
for the ``{\em classical\/}" wavefunctions
\begin{equation}
\hat{A}\varphi_n={\mathcal E}_n \varphi_n,\quad n=1,2,\ldots, ,
\end{equation}
in which the operator $\hat{A}$ is defined above (\ref{cleigen}).
Conversely one could start with
the above eigenvalue equation.
One defines the classical eigenfunctions as  its solutions
satisfying
certain regularity  conditions.
 Then 
the {\em quantum\/} eigenfunction
$\phi_n$ could be considered as 
 an $\hbar$-{\em deformation\/} of the {\em classical\/}
eigenfunction
$\varphi_n$. For the Calogero and Sutherland systems to be discussed below,
there is a one-to-one correspondence between the 
classical and quantum eigenfunctions.
For generic multi-particle quantum mechanical systems, 
the situation is less clear.

\bigskip

\subsection{Theorems}
\label{mainres}
The {\em classical\/}
eigenfunctions have the following remarkable properties:
\newtheorem{prop}{Proposition}[section]
\begin{prop}
\label{prop21}
The product of two {\em classical\/}
eigenfunctions $(\varphi_n,{\mathcal E}_n)$ and $(\varphi_m,{\mathcal E}_m)$
is again a {\em classical\/}
eigenfunction with the eigenvalue ${\mathcal E}_n+{\mathcal E}_m$,
\begin{equation}
-\sum_{j=1}^r{\partial W\over{\partial q_j}}{\partial
(\varphi_n\varphi_m)\over{\partial q_j}} =({\mathcal E}_n+{\mathcal E}_m)
\varphi_n\varphi_m.
\label{cleigen2}
\end{equation}
\end{prop}
\begin{prop} 
\label{prop22}
The classical eigenfunctions vanish at the equilibrium 
$\bar{q}$
\begin{equation}
\varphi_n(\bar{q})=0, \quad n=1,2,\ldots, .
\end{equation}
\end{prop}
\begin{prop} 
\label{prop23}
The derivatives of a classical eigenfunction  at the equilibrium  
$\bar{q}$ form an eigenvector of the Hessian matrix $\widetilde{W}$,
iff $\left.\nabla \varphi_n\right|_{\bar{q}}\neq0$
\begin{eqnarray}
-\widetilde{W}\cdot\left.\nabla \varphi_n\right|_{\bar{q}}
&=&{\mathcal E}_n\left.\nabla \varphi_n\right|_{\bar{q}},
\quad n=1,2,\ldots, .
\label{cleigen3}\\
&\mbox{\rm or}&\nonumber\\
-\sum_{j=1}^r\left.{\partial^2 W\over{\partial q_k \partial
q_j}}\right|_{\bar{q}}\left.{\partial
\varphi_n\over{\partial q_j}}\right|_{\bar{q}} &=&{\mathcal E}_n
\left.{\partial \varphi_n\over{\partial q_k}}
\right|_{\bar{q}},\quad n=1,2,\ldots, .
\label{cleigen4}
\end{eqnarray}
\end{prop}

Obviously the Hessian matrix $\widetilde{W}$ has at most $r$ different 
eigenvalues and eigenvectors.
The classical eigenfunctions $\{(\varphi_j,{\mathcal E}_j)\}$, $j=1,\ldots,r$ for
which
$\left.\nabla \varphi_j\right|_{\bar{q}}\neq0$ will be called ``{\em elementary
excitations\/}". At equilibrium, each corresponds to the {\em normal coordinate\/}
of the small oscillations with the eigen (angular) frequency ${\mathcal E}_j$.
The elementary
excitations are the generators of all the classical eigenfunctions. 
In other words, any classical eigenfunction can be expressed as
\begin{equation}
\varphi_1^{n_1}\cdots\varphi_r^{n_r},\quad {\mathcal E}=
n_1{\mathcal E}_1+\cdots+
n_r{\mathcal E}_r,\quad n_j\in{\mathbb Z}_+,
\label{linspec}
\end{equation}
or a linear combination thereof with the same eigenvalue ${\mathcal E}$.
The above type of classical eigenfunctions  are obviously non-elementary and they 
have zero gradient at equilibrium, for example,
$\left.\nabla(\varphi_j\varphi_k)\right|_{\bar{q}}=0$.
Because of this property, the representation of the elementary excitations
is not unique except for some lower members.

These results provide a basis of the analytical proof of the  observations
made in Corrigan-Sasaki  paper \cite{cs} on the correspondence/contrast
between  the {\em classical \/} and {\em quantum\/} integrability
in Calogero-Moser systems.
It should be mentioned that Perelomov's recent work 
\cite{perenew} asserts essentially
our
Proposition 2.3 for the special cases of the
quantum-classical eigenvalue correspondence of the Sutherland systems.

Throughout this section we have assumed that the prepotential $W$
is independent of the Planck's constant $\hbar$, for simplicity of the
presentation. The main content of this section is valid even if $W$ depends
on $\hbar$, so long as $\lim_{\hbar\to0}W=W_0$ is well-defined.
A celebrated example that $\lim_{\hbar\to0}W$ diverges is the hydrogen
atom, for which the classical equilibrium does not exist.
In this case the quantum-classical correspondence does not make sense
and the present formulation does not apply.

In the subsequent sections we will show many explicit examples of the  
{\em classical \/} and 
{\em quantum\/} eigenfunctions and their relationship.

\section{Root Systems and Calogero-Moser Dynamics}
\label{roots}
\setcounter{equation}{0}
A   Calogero-Moser system is a
multi-particle Hamiltonian dynamics associated with a root system \(\Delta\)
of rank \(r\).
This is a set of
vectors in \(\mathbf{R}^{r}\)  
invariant under reflections
in the hyperplane perpendicular to each
vector in $\Delta$:
\begin{equation}
  \Delta\ni s_{\alpha}(\beta)=\beta-(\alpha^{\vee}\cdot\beta)\alpha,
   \quad \alpha^{\vee}={2\alpha\over{\alpha^2}},
\quad \alpha, \beta\in\Delta.\label{a1}
\end{equation}
The set of reflections $\{s_{\alpha},\,\alpha\in\Delta\}$ generates a
finite reflection group $G_{\Delta}$, known as a Coxeter (or Weyl) group.

A   Calogero-Moser system is integrable both at the classical 
and quantum levels for
various choices of the long range interaction potentials;
rational ($1/q^2$), rational with a harmonic confining potential,
trigonometric ($1/\sin^2q$), hyperbolic ($1/\sinh^2q$) and elliptic 
with the Weierstrass function ($\wp(q)$) potential.
In the rest of this paper we will discuss 
the rational case (with a harmonic confining potential)
under the name of   Calogero system \cite{Cal} and the trigonometric
potential  case to be called  Sutherland system \cite{Cal}.
Both quantum Hamiltonians  have an infinite number of discrete eigenvalues.
The prepotentials are
\begin{equation}
\mbox{Calogero}:\qquad W=W_R-{\omega\over2}q^2,\quad 
W_R=\sum_{\rho\in\Delta_+}g_{\rho}\log\rho\cdot q,
\label{CalW}
\end{equation}
in which $W_R$ is the prepotential of the theory without the harmonic
confining potential and 
\begin{equation}
\mbox{Sutherland}:\qquad W=\sum_{\rho\in\Delta_+}g_{\rho}\log\sin(\rho\cdot
q).\hspace*{3.3cm}
\label{SutW}
\end{equation}
%
In these formulae, $\Delta_+$ is the set of positive roots and
 \(g_{\rho}\) are  real {\em positive\/} coupling constants
which are defined on orbits of the corresponding
Coxeter group, {\it i.e.} they are
identical for roots in the same orbit. 
For crystallographic root
systems there is one
coupling constant
\(g_{\rho}=g\) for all roots in simply-laced models,
and  there are two {\em independent\/} coupling constants,
\(g_{\rho}=g_L\) for long roots and \(g_{\rho}=g_S\) for
short roots in non-simply laced models.
We will give the explicit forms of $W$ in later sections.
Throughout this paper we put the scale factor in the trigonometric 
functions to
unity for simplicity; instead of the general form
$a^2/\sin^2a(\rho\cdot q)$, we use $1/\sin^2(\rho\cdot q)$.
We also adopt the convention that 
long roots have squared length two, $\rho_L^2=2$,
unless otherwise stated.
These prepotentials determine the potentials:
\begin{eqnarray}
V=\left\{\begin{array}{l}
\begin{displaystyle}
{\omega^2\over2}q^2 +{1\over2}\sum_{\rho\in\Delta_+}
   {g_{\rho}(g_{\rho}-\hbar) \rho^{2}\over{(\rho\cdot q)^2}}-E_0,
\qquad \mbox{Calogero},
\end{displaystyle}
\\[20pt]
\begin{displaystyle}
{1\over2}\sum_{\rho\in\Delta_+}
   {g_{\rho}(g_{\rho}-\hbar) \rho^{2}\over{\sin^2(\rho\cdot q)}}-E_0,
\qquad\qquad\quad\ \mbox{Sutherland}.
\end{displaystyle}
\end{array}
\right.
\label{fullpot}
\end{eqnarray}

The Sutherland systems  are integrable,
both at the classical and quantum levels,
for the crystallographic root systems, 
 that is
those associated with simple Lie
algebras: \{\(A_{r},\,r\ge 1\}\)%
\footnote{For  $A_r$ models, it is customary to
introduce one more degree of freedom,
$q_{r+1}$ and $p_{r+1}$ and embed
all of the roots in ${\bf R}^{r+1}$.
\label{embedding}}
, \(\{B_{r},\,r\ge 2\}\), \(\{C_{r},\,r\ge
2\}\),
\(\{D_{r},\,r\ge 4\}\), \(E_{6}\), \(E_{7}\), \(E_{8}\), \(F_{4}\) and
\(G_{2}\) and the so-called \(\{BC_{r},\,r\ge 2\}\).
On the other hand, the Calogero systems  are
integrable for any root systems, crystallographic and
non-crystallographic. The latter are \(H_{3}\), \(H_{4}\), 
and  \(\{I_{2}(m),\,m\ge 4\}\), the dihedral group of order \(2m\).

\bigskip
The prepotential $W$ (\ref{CalW}), (\ref{SutW}), and hence the (classical)
potential
$V$ (\ref{fullpot}) and the Hamiltonian are Coxeter (Weyl) invariant: 
\begin{eqnarray}
W(s_\rho(q))&=&W(q),\quad V(s_\rho(q))=V(q),\quad
V_C(s_\rho(q))=V_C(q),\quad \forall
\rho\in
\Delta,\nonumber\\
H(s_\rho(p),s_\rho(q))&=&H(p,q),\quad
\hat{H}(s_\rho(p),s_\rho(q))=\hat{H}(p,q),\quad
\hat{A}(s_\rho(q))=\hat{A}(q),
\end{eqnarray}
which is the symmetry of the entire Calogero-Moser systems.
This results in the fact that the ground state $\psi_0$ and all
the other eigenfunctions are are Coxeter (Weyl) invariant \cite{kps}: 
\begin{equation}
\psi_0(s_\rho(q))=\psi_0(q),\quad \psi_n(s_\rho(q))=\psi_n(q),\quad
\phi_n(s_\rho(q))=\phi_n(q),\quad \varphi_n(s_\rho(q))=\varphi_n(q).
\end{equation}

The quantum Calogero and Sutherland systems are not only integrable but also
{\em exactly solvable\/} \cite{kps}, that is, the similarity transformed
Hamiltonians (\ref{htilform}) are {\em lower triangular\/} in certain basis
of the Hilbert space.
The eigenvalues can be read off easily from the diagonal matrix elements of
$\hat{H}$. The exact eigenvalues of the excited states in the Calogero
system are an integer multiple of the oscillator quantum $\omega\hbar$:
\begin{equation}
E_{\vec{n}}=\omega\hbar\sum_{j=1}^r n_jf_j,\quad  
\quad n_j\in{\mathbb Z}_+.
\end{equation}
Here $\vec{n}=(n_1,\ldots,n_r)$ are non-negative {\em quantum numbers\/}
and $f_j=1+e_j$ and the integers
$\{e_j\}$,
$j=1,\ldots,r$ are called the {\em exponents\/}
 of the root system $\Delta$:
\begin{center}
   	\begin{tabular}{||c|l||c|l||}
      	\hline
      	 \(\Delta\)& \(f_j=1+e_j\) &\(\Delta\)& \(f_j=1+e_j\)\\
      	\hline
      	\(A_r\) & \(2,3,4,\ldots,r+1\) & \(E_8\) & \(2,8,12,14,18,20,24,30\) \\
      	\hline
      	\(B_r\) & \(2,4,6,\ldots,2r\) & \(F_4\) & \(2,6,8,12\) \\
      	\hline
      	\(C_r\) & \(2,4,6,\ldots,2r\) & \(G_2\) & \(2,6\) \\
      	\hline
      \(D_r\) & \(2,4,\ldots,2r-2;r\) & \(I_2(m)\) & \(2,m\) \\
      \hline
      \(E_6\) & \(2,5,6,8,9,12\) & \(H_3\) & \(2,6,10\) \\
      \hline
      \(E_7\) & \(2,6,8,10,12,14,18\) & \(H_4\) & \(2,12,20,30\) \\
      \hline
   	\end{tabular}\\
	\bigskip
	Table I: The degrees \(f_j\) at an elementary excitation exists.
\end{center}
The coupling constant(s) $g_\rho$ of the rational $1/q^2$
potentials shifts {\em only\/} the ground state energy $E_0$
in (\ref{fullpot}):
\begin{equation}
E_0=\omega\hbar r/2+\omega\sum_{\rho\in{\Delta_+}}g_\rho
\label{E0def}
\end{equation}
in which the first term is the zero-point energy of the
oscillators.
For a given non-negative integer $N$, let $\mathcal{P}(N)$ be the 
number of different solutions of 
\begin{equation}
N=\sum_{j=1}^r n_jf_j, \quad n_j\in{\mathbb Z}_+.
\label{partitions}
\end{equation}
Then the energy eigenvalue $E=\omega\hbar N+E_0$ has the degeneracy
$\mathcal{P}(N)$.

The exact eigenvalues of the excited states in the Sutherland
\cite{kps,HeOp}
system are specified by the {\em dominant highest weight\/}
$\lambda_{\vec{n}}$:
\begin{eqnarray}
E_{\vec{n}}&=&
2\hbar^2\lambda_{\vec{n}}^2+4\hbar\lambda_{\vec{n}}\cdot\varrho,
\label{Sutspec}\\
\lambda_{\vec{n}}&=&\sum_{j=1}^rn_j\lambda_j,\quad 
n_j\in{\mathbb Z}_+,
\label{domweight}\\
\varrho&=&{1\over2}\sum_{\rho\in\Delta_+}g_{\rho}\rho,\quad
\delta={1\over2}\sum_{\rho\in\Delta_+}\rho.
\label{weylvec}
\end{eqnarray}
Here $\vec{n}=(n_1,\ldots,n_r)$ are non-negative {\em quantum numbers\/},
$\lambda_j$, $j=1,\ldots,r$ are the {\em fundamental weights\/}
and $\delta$ and $\varrho$  are called the {\em Weyl vector\/}
 and a {\em deformed
Weyl vector\/}. The ground state energy $E_0$ in (\ref{fullpot}) is solely
determined by
$\varrho$:
\begin{equation}
E_0=2\varrho^2.
\end{equation}

For the general discussion of quantum Calogero and Sutherland systems
for any root system
along the present line of arguments,
 the quantum integrability, Lax pairs, quantum eigenfunctions, 
creation-annihilation operators etc, we refer to \cite{kps}.
A rather different approach by Heckman and Opdam
\cite{HeOp} to Calogero-Moser models with degenerate potentials
based on any root system should also be mentioned in this connection.
The eigenfunctions of the Sutherland systems are sometimes called 
Heckman-Opdam's Jacobi polynomials.
Those for  the $A$ series  are known as the Jack
polynomials \cite{Stan}.

In the following two sections, we will show the classical and quantum
eigenfunctions of the elementary excitations in Calogero systems
(section \ref{rationals}) and in Sutherland systems (section \ref{trigs}).
For brevity  and clarity of the presentation, we present the
eigenfunctions of the `{\em reduced
theory\/}' in which most of the coupling constants are put to unity. 
To be
more precise, for simply-laced theories  ($A$, $D$, $E$, $H$ and $I_2($odd)) 
we put the coupling constant unity, $g=1$. For non simply-laced
theories ($B$, $C$, $F_4$, $G_2$ and $I_2($even)) we put the
coupling constant for long roots unity $g_L=1$ and keep
the coupling constant for short roots intact, $g_S=\gamma$.
The angular frequency of the harmonic confining potential is also put to
unity,
$\omega=1$.

Let us introduce elementary symmetric polynomials as useful
ingredients for expressing the eigenfunctions.
The degree $k$ elementary symmetric polynomial in $r$ variables,
$\{t_1,t_2,\ldots,t_r\}$, $S_k(\{t_j\})$ is defined by the expansion of  a
generating function 
\begin{equation}
G(x;\{t_j\})=\prod_{j=1}^r(x+t_j)=\sum_{k=0}^r S_k(\{t_j\})x^{r-k}.
\label{genfun}
\end{equation}

\section{Classical \& Quantum Eigenfunctions of the Calogero Systems}
\label{rationals}
\setcounter{equation}{0}

The basis  of the quantum eigenfunctions $\{\phi_n\}$ of the
Calogero system is
the Coxeter (Weyl) invariant polynomials in the coordinates $\{q_j\}$.
In order to express the eigenfunctions in a closed form, let us introduce
the similarity transformed Hamiltonian $\hat{H}_R$ {\em without the harmonic
confining potential\/}:
\begin{eqnarray}
\hat{H}_R&=&\hat{H}-\omega\hbar{\mathcal D},\qquad\qquad\
{\mathcal D}=q\cdot\nabla=\sum_{j=1}^r q_j{\partial\over{\partial q_j}},\\
&=&-{\hbar^2\over2}\triangle +\hbar\hat{A}_R,\qquad
\hat{A}_R=-\nabla W_R\cdot\nabla=-\sum_{j=1}^r{\partial W_R\over{\partial
q_j}}{\partial\over{\partial q_j}},
\label{ARdef}
\end{eqnarray}
in which $\mathcal{D}$ is the Euler derivative measuring the degree
of a monomial.
The Hamiltonian $\hat{H}_R$  maps a Coxeter invariant polynomial to another
{\em with degree two less\/}
\begin{equation}
[\mathcal{D}, \hat{H}_R]=-2\hat{H}_R,
\end{equation}
which implies for an arbitrary parameter $\kappa\in\mathbb{C}$
\[
(\mathcal{D}+2\kappa
\hat{H}_R)\exp[\kappa\hat{H}_R]=\exp[\kappa\hat{H}_R]\mathcal{D}.
\]
The {\em lower triangularity} of the Hamiltonian $\hat{H}_R$ means  that the
exponential operator contains only finite powers of $\hat{H}_R$, up to
$[N/2]$, when applied to a Coxeter invariant polynomial of degree $N$. 
 By multiplying $\omega\hbar$ on both sides and choosing
$\kappa=1/2\omega\hbar$, we obtain
\begin{equation}
\hat{H}\exp[\hat{H}_R/2\omega\hbar]=
\exp[\hat{H}_R/2\omega\hbar]\omega\hbar\mathcal{D}.
\end{equation}
Thus we arrive at a formula of an eigenfunction of $\hat{H}$ with the
eigenvalue $\omega\hbar N$ ($N$ being a non-negative integer), starting from
an arbitrary {\em homogeneous\/} Coxeter invariant polynomial $\Phi_N(q)$
of degree $N$:
\begin{eqnarray}
\hat{H}\phi_N(q)&=&\omega\hbar N \phi_N(q),
\label{qeigform1}\\
\phi_N(q)&\equiv&\exp[\hat{H}_R/2\omega\hbar]\Phi_N(q),\quad
\mathcal{D}\Phi_N(q)=N\Phi_N(q).
\label{qeigform2}
\end{eqnarray}
A similar formula was derived in \cite{SogoGrPan}
 for the theories based on 
the $A$-series of root systems. 
There are $\mathcal{P}(N)$ (\ref{partitions}) linearly independent
Coxeter invariant homogeneous polynomials, which is  
equal to the degeneracy of the
eigenspace of $E=\omega\hbar N$. 
Among them there are special eigenfunctions which are linear combinations of
the Coxeter invariant {\em homogeneous\/} polynomials such that they are
annihilated by $\hat{H}_R$:
\begin{equation}
\hat{H}_R\Phi_N=0\quad \Longrightarrow 
\quad\hat{H}\Phi_N(q)=\omega\hbar N \Phi_N(q).
\end{equation}
The number of homogeneous eigenfunctions is $\mathcal{P}(N)-\mathcal{P}(N-2)$,
which is much less  than the total dimensionality 
of the eigenspace, $\mathcal{P}(N)$.

The simplest class of quantum eigenfunctions depends only on 
$q^2=\sum_{j}q_j^2$,
(\ref{qpdefs}):
\begin{eqnarray}
\hat{A}_R(q^2)^n&=&-2n\sum_{\rho\in\Delta_+}g_\rho (q^2)^{n-1}, 
\quad
\triangle (q^2)^n=4n(r/2+n-1)(q^2)^{n-1},\nonumber\\
\hat{H}_R (q^2)^n&=&-2n\hbar\left({r\over2}\hbar +\sum g_\rho +\hbar(n-1)\right)
(q^2)^{n-1},\nonumber\\
\hat{H}_R \left({\omega\over\hbar}q^2\right)^n&=&-2n\omega\hbar\left({r\over2}
+{1\over\hbar}\sum g_\rho +(n-1)\right)
\left({\omega\over\hbar}q^2\right)^{n-1}.
\end{eqnarray}
If we define $x\equiv\omega q^2/\hbar$ and
$\alpha\equiv r/2+\sum_{\rho}g_\rho/\hbar-1=E_0/\hbar\omega-1$,
 (with $E_0$ defined in
(\ref{E0def})),
we obtain the associated Laguerre polynomial in $x$ as the quantum eigenfunction
\begin{equation}
\exp\left[{\hat{H}_R\over{2\omega\hbar}}\right](-1)^n{x^n\over{n!}}=
\sum_{j=0}^n(-1)^j{\alpha+n\choose n-j}{x^j\over{j!}}=L^{(\alpha)}_n(x).
\label{laguerrepoly}
\end{equation}
This class of universal eigenfunctions is known from the early days of
Calogero systems \cite{Gamb,kps}.
It is easy to verify Proposition \ref{prop22} and 
\ref{prop23} for this eigenfunction,
since the classical limit  is
\begin{equation}
\lim_{\hbar\to0}\hbar^n L^{(\alpha)}_n(x)
=(-)^n{\omega^n\over{n!}}(q^2-\bar{q}^2)^n,
\end{equation}
in which $\omega\bar{q}^2=\sum_{\rho\in\Delta_+}g_\rho$ \cite{cs}.

The classical counterpart of the above general result (\ref{qeigform1}),
(\ref{qeigform2}) is simply obtained  as the $\hbar\to0$ limit:
\begin{eqnarray}
\hat{A}\varphi_N(q)&=&\omega N \varphi_N(q),\\
\varphi_N(q)&\equiv&\exp[\hat{A}_R/2\omega]\Phi_N(q),\quad
\mathcal{D}\Phi_N(q)=N\Phi_N(q).
\end{eqnarray}
Since the operator $\hat{A}_R$ ($\hat{A}$) satisfies the Leibnitz rule
$\hat{A}_R(fg)=(\hat{A}_Rf)g+f(\hat{A}_Rg)$, we obtain 
corresponding to Proposition 2.1 (\ref{cleigen2})
\begin{equation}
\exp[\hat{A}_R/2\omega]\Phi_N(q)\Phi_M(q)=
\{\exp[\hat{A}_R/2\omega]\Phi_N(q)\}\,
\{\exp[\hat{A}_R/2\omega]\Phi_M(q)\}.\label{mulkform}
\end{equation}
The classical eigenfunctions of the elementary
excitations are the generators of all the classical eigenfunctions.
The  quantum eigenfunctions of the elementary
excitations play a less prominent role.
The product of two quantum eigenfunctions is no longer a
quantum eigenfunction, since the Laplacian $\triangle$ and thus the
Hamiltonian $\hat{H}$ do not enjoy the Leibnitz rule.
Here we will show explicitly the classical and quantum eigenfunctions of
the elementary excitations for the Calogero systems.
The knowledge of the classical equilibrium and
the eigenvectors of the Hessian matrix $\widetilde{W}$ (\ref{VWmat}) helps to
determine the classical eigenfunctions.

\subsection{A-Series}
\label{calA}
Calogero and collaborators discussed the classical equilibrium problem 
of the $A_r$ Calogero system 
about quarter of a century ago 
\cite{calmat,calpere,ahmcal,OP1}.
A modern version in terms of the prepotential was developed by Corrigan-Sasaki
\cite{cs}. 
Following the usual convention we embed the root vectors  in ${\bf R}^{r+1}$
as:
\begin{equation}
A_r=\{{\bf e}_j-{\bf e}_k,\
j,k=1,\ldots,r+1|{\bf e}_j\in{\bf R}^{r+1}, {\bf e}_j\cdot
{\bf e}_k=\delta_{jk}\}.
\end{equation}
The prepotential  for the full and {\em reduced\/} theory read
\begin{equation}
W=g\sum_{j<k}^{r+1}\log(q_j-q_k)-{\omega\over2}q^2,\quad
W=\sum_{j<k}^{r+1}\log(q_j-q_k)-{1\over2}q^2.
\end{equation}
We discuss the reduced theory for simplicity and brevity.
The equations (\ref{Vmin}) determining the maximum of the ground state
wavefunction $\psi_0$ read
\begin{equation}
\sum_{k\ne j}^{r+1}{1\over{\bar{q}_j-\bar{q}_k}}=
\bar{q}_j,
\quad j=1,\ldots,r+1.
\label{herroots}
\end{equation}
These determine $\{\bar{q}_j\}$,
$j=1,\ldots,r+1$ to be the zeros of the Hermite  polynomial $H_{r+1}(x)$
\cite{calmat,szego,cs},
\[
H_{r+1}(\bar{q}_j)=0.
\]
The Hessian $-\widetilde{W}$ has eigenvalues $\{1,2,\ldots,r+1\}$,
which are exactly the quantum eigenvalues (divided by $\omega\hbar$) of the
elementary excitations   listed
in Table I. (The lowest eigenvalue 1 belongs to the center of mass degree
of freedom which is completely decoupled from the other modes.)

Here are our new results on the classical and quantum eigenfunctions.
The $k$-th eigenvector of $\widetilde{W}$ has a simple form \cite{perepoin}
\begin{equation}
v_k=(P_k(\bar{q}_1),\ldots,P_k(\bar{q}_{r+1})),
\quad k=0,\ldots,r
\label{vkPform}
\end{equation}
in which $P_k(x)$ is a  polynomial of degree $k$ of a single 
variable. They  obey the
following three term recursion relation:
\begin{equation}
P_k(x)=x\,P_{k-1}(x) +{k-r-2\over{2}}P_{k-2}(x),
\quad P_0(x)=1,\quad P_1(x)=x.
\label{aPrec}
\end{equation}
The orthogonality relations of the eigenvectors $\{v_k\}$ read simply as
\begin{equation}
v_j\cdot v_k=0 \Longleftrightarrow 
\sum_{l=1}^{r+1}P_j(\bar{q}_l)P_k(\bar{q}_l)=0,\quad j\neq k=0,1,\ldots, r.
\end{equation}
These are `orthogonal polynomials of a discrete variable'
\cite{szego,bateman}. 
In the present case, the discrete variable is obviously the zeros of
the Hermite polynomial.

A simple representation of the elementary excitations is provided by
the elementary symmetric polynomials in $\{q_j\}$, $S_k(\{q_j\})$,
$k=0,1,\ldots,r+1$ (\ref{genfun})
which are obviously Weyl invariant.
By applying the operator $\hat{A}_R$ (\ref{ARdef}) 
on the generating function (\ref{genfun}) 
\begin{equation}
G(x;\{q_j\})=\prod_{j=1}^{r+1} (x+q_j)
\end{equation}
and noting $\partial_{q_i}G=G/(x+q_i)$ and
$\partial_x^2G=\sum_{i,j}G/(x+q_i)(x+q_j)$, we obtain
\begin{equation}
\hat{A}_RG=\frac{1}{2}\partial_x^2G
\label{acalgeneq}
\end{equation}
through partial fraction
decomposition. 
This leads to
\begin{eqnarray}
\hat{A}S_k(\{q_j\})&=&kS_k+{(r+3-k)(r+2-k)\over2}S_{k-2}(\{q_j\}).
\label{aASk}
\end{eqnarray}
We obtain the classical
eigenfunctions for the elementary excitations:
\begin{equation}
\hat{A}\varphi_k(q)=k\varphi_k(q),\quad
\varphi_k(q)=\exp[\hat{A}_R/2]S_k(\{q_j\})=\sum_{l=0}^{[k/2]}{(r+1-k+2l)!
\over{(r+1-k)!4^ll!}}S_{k-2l}(\{q_j\}).
\end{equation}
Since  $\varphi_k$ are harmonic polynomials
\begin{equation}
\triangle S_k(\{q_j\})=0 \Longrightarrow \triangle\varphi_k=0, \qquad
k=0,1,\ldots,r+1,
\label{aLapSk}
\end{equation}
they are at the same time  quantum eigenfunctions:
\begin{equation}
\hat{H}\varphi_k(q)=k\hbar\varphi_k(q),\qquad k=1,\ldots,r+1.
\end{equation}
With some calculation one can verify Propositions 2.2--2.3, that is
$\varphi_k(\bar{q})=0$ and its derivative gives the above
function $P_k$, $\partial/\partial
q_j\varphi_k(\bar{q})=(-)^kP_{k-1}(\bar{q}_j)$.
Indeed, as $\partial_{q_j}S_k(\{q\})=S_{k-1}(\{q\})-q_j
S_{k-2}(\{q\})+\cdots+(-q_j)^{k-1}$, one finds that
\begin{eqnarray}
\partial_{q_j}\varphi_k(\bar q)&=&(-)^{k-1}\left[\bar
q_j^{k-1}+\left(S_2(\{\bar q\})+(r+3-k)(r+2-k)/4\right)\bar
q_j^{k-3}+\ldots\right]\nonumber\\
&\equiv& (-)^{k-1}P_{k-1}(\bar q_j). 
\label{recdetails}
\end{eqnarray}
The
polynomial $P_k(x)$ of degree $k$ (with $0\leq k\leq r$) cannot 
vanish in all the $r+1$ points $\bar q_j$ (with $1\leq j\leq r+1$).
Hence $\nabla \varphi_k(\bar q)\neq 0$ and it follows that
$\varphi_k(q)$ is indeed an elementary excitation.
As we now know that the expressions (\ref{vkPform}) are eigenvectors corresponding to 
different eigenvalues of the matrix  $-\tilde W$, 
it follows that the $P_k(x)$ are orthogonal polynomials of a discrete variable. 
Hence they obey a three term recurrence relation of type 
$P_k(x)=(A_k+B_k x)P_{k-1}(x)+C_k P_{k-2}(x)$. 
The coefficients of this recurrence are obtained from the definition 
(\ref{recdetails}) for different $k$, {\em i.e.\/} $A_k=0$, $B_k=1$, $C_k=(k-r-2)/2$.
\subsection{B-Series}
\label{calB}
Let us note that the rational $C_r$ and $BC_r$ systems are identical with
the $B_r$ system. 
The root vectors of $B_r$ are expressed neatly in terms of an
orthonormal basis of  ${\bf
R}^{r}$ as:
\begin{equation}
B_r=\{\pm{\bf e}_j \pm{\bf e}_k,\quad \pm {\bf e}_j,\quad 
j,k=1,\ldots,r|{\bf e}_j\in{\bf R}^{r}, {\bf e}_j\cdot
{\bf e}_k=\delta_{jk}\}.
\label{brroots}
\end{equation}
The prepotential for the full and {\em reduced\/} theory read
\begin{equation}
W=g_L\sum_{j<k}^{r}\log(q_j^2-q_k^2)+g_S\sum_{j=1}^{r}\log
q_j-{\omega\over2}q^2,\quad
W=\sum_{j<k}^{r}\log(q_j^2-q_k^2)+\gamma\sum_{j=1}^{r}\log
q_j-{1\over2}q^2.
\end{equation}
We discuss the reduced theory.
 Assuming $\bar{q}_j\ne0$, the equations (\ref{Vmin}) 
determining the maximum of the ground state
wavefunction $\psi_0$ read
\begin{equation}
\sum_{k\ne
j}^{r}{1\over{\bar{q}_j^2-\bar{q}_k^2}}
+{\gamma/2\over{\bar{q}_j^2}}={1\over{2}},
\quad j=1,\ldots,r,
\label{lagroots}
\end{equation}
and determine $\{\bar{q}_j^2\}$, $j=1,\ldots,r$, as the
zeros of the associated Laguerre  polynomial $L_{r}^{(\gamma-1)}(x)$,
\cite{szego,OP1,cs}, 
\(
L_r^{\gamma-1}(\bar{q}^2_j)=0
\).
The Hessian $-\widetilde{W}$ has eigenvalues $\{2,4,6,\ldots,2r\}$,
which are exactly the quantum eigenvalues (divided by $\omega\hbar$) of the
elementary excitations   listed
in Table I. 

The new results on the classical and quantum eigenfunctions are as follows.
The $k$-th eigenvector of $\widetilde{W}$ has a simple
form
\begin{equation}
v_{k-1}=(\bar{q}_1P_{k-1}(\bar{q}^2_1),\ldots,\bar{q}_{r}P_{k-1}(\bar{q}^2_{r})),
\quad k=1,\ldots,r,\quad \bar{q}_l>0,
\label{brosc}
\end{equation}
in which the polynomials $\{P_k(x)\}$ obey the following
three term recursion relation:
\begin{eqnarray}
P_k(x)&=&(\vTm x-2(r-k)-\gamma)\,P_{k-1}(x) -(k-r-\gamma)(k-r-1)P_{k-2}(x),
\label{bPrec1}\\
P_0(x)&=&1,\quad P_1(x)=x-2(r-1)-\gamma.
\label{bPrec}
\end{eqnarray}
The orthogonality relations of the eigenvectors $\{v_k\}$ again correspond to
those of orthogonal polynomials of a discrete variable:
\begin{equation}
v_j\cdot v_k=0 \Longleftrightarrow 
\sum_{l=1}^{r}\bar{q}^2_lP_j(\bar{q}^2_l)P_k(\bar{q}^2_l)=0,\quad j\neq
k=0,\ldots, r-1.
\end{equation}

\bigskip
A simple representation of the elementary excitations is provided by
the elementary symmetric polynomials in $\{q_j^2\}$, $S_k(\{q_j^2\})$,
$k=0,1,\ldots,r$,
which are obviously Weyl invariant.
By applying the operator $\hat{A}_R$ (\ref{ARdef}) and the
Laplacian $\triangle$ on the generating function (\ref{genfun}) 
$G(x;\{q_j^2\})=\prod_{j=1}^r (x+q_j^2)$, we obtain
\begin{equation}
\hat{A}_RG=-2x\partial^2_xG-2\gamma\partial_xG\quad \mbox{and}\quad 
\triangle G=2\partial_xG.
\label{bcalgeneq}
\end{equation}
For the former formula, as in the $A$-series case,  the partial fraction
decomposition is used.
These mean 
\begin{eqnarray}
\hat{A}S_k(\{q_j^2\})&=&2kS_k(\{q_j^2\})+2(r-k+1)(k-r-\gamma)S_{k-1}(\{q_j^2\}),
\label{brAonSk}\\
\triangle S_k(\{q_j^2\})&=&2(r-k+1)S_{k-1}(\{q_j^2\}), \qquad k=1,\ldots,r,
\label{brtrionSk}
\end{eqnarray}
from which we obtain the classical eigenfunctions for the elementary
excitations:
\begin{eqnarray}
\hat{A}\varphi_k(q)&=&2k\varphi_k(q),\qquad k=1,\ldots,r,
\label{brclas}\\
\varphi_k(q)&=&\exp[\hat{A}_R/2]S_k(\{q_j^2\})=\sum_{l=0}^{k}(-)^l
{r+l-k\choose l}{\Gamma(r+l-k+\gamma)
\over{\Gamma(r-k+\gamma)}}S_{k-l}(\{q_j^2\}).
\nonumber
\end{eqnarray}
The corresponding quantum eigenfunctions have  very similar forms, since the
action of
$-{\hbar/2}\triangle +\hat{A}$ on $S_k$ (\ref{brAonSk}), (\ref{brtrionSk}) is
the same as that of 
$\hat{A}$ with
$\gamma$ replaced by $\gamma+\hbar/2$:
\begin{eqnarray}
\hat{H}\phi_k(q)&=&2k\hbar\phi_k(q),\qquad k=1,\ldots,r,
\label{brquant}\\
\phi_k(q)&=&\exp[\hat{H}_R/2\hbar]S_k(\{q_j^2\})=\sum_{l=0}^{k}(-)^l
{r+l-k\choose l}{\Gamma(r+l-k+\gamma+\hbar/2)
\over{\Gamma(r-k+\gamma+\hbar/2)}}S_{k-l}(\{q_j^2\}).
\nonumber
\end{eqnarray}
The correspondence between the (classical) eigenfunctions
(\ref{brclas}), (\ref{brquant}) and the classical eigenvectors
(\ref{brosc}) (and (\ref{bPrec1}), (\ref{bPrec})) is established in an identical
manner as in the
$A$-case.
\subsection{D-Series}
\label{calD}
The root vectors of $D_r$ are:
\begin{equation}
D_r=\{\pm{\bf e}_j \pm{\bf e}_k,\
j,k=1,\ldots,r|{\bf e}_j\in{\bf R}^{r}, {\bf e}_j\cdot
{\bf e}_k=\delta_{jk}\}.
\label{drroots}
\end{equation}
The prepotential for the full and {\em reduced\/} theory read
\begin{equation}
W=g\sum_{j<k}^{r}\log(q_j^2-q_k^2)
-{\omega\over2}q^2,\quad
W=\sum_{j<k}^{r}\log(q_j^2-q_k^2)
-{1\over2}q^2.
\end{equation}
We discuss the reduced theory.
The equilibrium position
$\bar{q}_j^2$ are the zeros of the Laguerre polynomial $L_r^{-1}(x)$,
that is one of $\{\bar{q}_j\}$'s is vanishing \cite{cs}.
Let us choose $\bar{q}_1=0$.

The Hessian $-\widetilde{W}$ has eigenvalues $\{2,4,6,\ldots,2(r-1),r\}$,
which are exactly the quantum eigenvalues (divided by $\omega\hbar$) of the
elementary excitations   listed
in Table I. 
The $k$-th eigenvector of $\widetilde{W}$ has a simple form
\begin{eqnarray}
v_{k}&=&(0,\bar{q}_2P_{k-1}(\bar{q}^2_2),
\ldots,\bar{q}_{r}P_{k-1}(\bar{q}^2_{r})),
\quad k=1,\ldots,r-1,\quad \bar{q}_l>0,\\
v_r&=&(1,0,\ldots,0),
\end{eqnarray}
in which the polynomials $\{P_k(x)\}$ obey the following
three term recursion relation:
\begin{eqnarray}
P_k(x)&=&(\vTm x-2(r-k))\,P_{k-1}(x) -(k-r)(k-r-1)P_{k-2}(x),\\
 P_0(x)&=&1,\quad P_1(x)=x-2(r-1),
\label{dPrec}
\end{eqnarray}
which is a special case of that of $B_r$ with $\gamma=0$.
The classical and quantum eigenfunctions $\varphi_1$, \ldots, $\varphi_{r-1}$
and $\phi_1$, \ldots, $\phi_{r-1}$ have the same form as in $B_r$ case
(\ref{brclas}), (\ref{brquant}) with
$\gamma=0$.
The special elementary excitation of the $D_r$ theory belongs to the $r$-th
eigenvalue. The classical eigenfunction is the same as the quantum one
\begin{eqnarray}
\varphi_r(q)&=&q_1q_2\cdots q_r,\quad \hat{A}\varphi_r=r\varphi_r,
\quad \triangle\varphi_r=0\Rightarrow \hat{H}\varphi_r=r\hbar\varphi_r,\\
\varphi_r(\bar{q})&=&0,\quad
\left.\vTmm\nabla\varphi_r(q)\right|_{\bar{q}}\propto(1,0,\ldots,0).
\end{eqnarray}
\subsection{E-Series}
\label{calE}
For the $E$-series of root systems we consider only the reduced theory, that is
$\omega=g=1$. The roots are normalised $\rho^2=2$, 
for all root systems $E_6, E_7$ and
$E_8$. 
\subsubsection{$E_6$}
\label{calE6}
The generating function is defined in terms 
of the weights belonging to the {\bf 27}
dimensional representation:
\begin{equation}
G(x;\mathbf{27})=\prod_{\mu\in\mathbf{27}}(x+\mu\cdot q)
=\sum_{k=0}^{27} S_k x^{27-k}.
\end{equation}
 These are {\em minimal weights\/}
having the properties $\rho\cdot\mu=\pm1,0$ for $\forall\rho\in E_6$ and
\begin{equation}
\mu^2=4/3,\quad \mu\cdot\nu=\left\{
\begin{array}{r}
1/3,\\
-2/3,
\end{array}
\right.\quad \mu\neq \nu \in\mathbf{27}.
\end{equation}
By applying the operator $\hat{A}_R$ (\ref{ARdef}) 
and using the above properties of
the weight vectors, we obtain
\begin{equation}
\hat{A}_RG={1\over2}\triangle G+{1\over3}\partial_x^2 G,
\label{e6calgeneq}
\end{equation}
or in terms of the $S_k$,
\begin{equation}
\hat{A}_RS_k={1\over2}\triangle S_k+{1\over3}(29-k)(28-k)S_{k-2}.
\end{equation}
Some lower members of $S_k$, which depend on $\{q_1,\ldots,q_6\}$  are
\begin{equation}
S_0=1,\ S_1=0,\ S_2=-3q^2,\ S_3=0, \ S_4={15\over4}(q^2)^2,\ldots.
\end{equation}
The elementary excitations are for $N=$2, 5, 6, 8, 9, 12, (see Table I) for which
$\{S_N\}$'s are functionally independent.  
The corresponding quantum eigenfunctions
are:
\begin{eqnarray}
\phi_2&=&q^2-3(12+\hbar),\\
\phi_5&=&S_5,\\
\phi_6&=&S_6+{(1110+155\hbar)/4}\left[(q^2)^2-4(9+\hbar)(q^2-(12+\hbar))\right],\\
\phi_8&=&S_8+(57+13\hbar)S_6-{15/2}(1-\hbar)(q^2)^3\\
&&\hspace*{2mm}+{5(6+\hbar)(2217+313\hbar)/24}\left[
3(q^2)^3-8(9+\hbar)q^2+6(9+\hbar)(12+\hbar)\right],\\
\phi_9&=&S_9-{35/6}(15+4\hbar) S_5\times[q^2-2(9+2\hbar)],\\[6pt]
\phi_{12}&=&
S_{12} +\frac{\left(1725534 + 1146267\hbar + 188608\hbar^2\right)%
      }{6144\left(3 + \hbar \right)%
      \left( 36 + 11\hbar \right)}(q^2)^6 +
   \frac{23}{576} S_5^2q^2+
   \frac{101\,}{192}S_6^2\nonumber\\
&&+
   \frac{\left( 222939 + 146112\hbar + 23798\hbar^2 \right) \,
      }{960\,\left( 3 + \hbar \right) \,
      \left( 36 + 11\hbar \right)}(q^2)^3S_6  -
   \frac{\left(3216 + 937\hbar \right) \,
      }{1920\,\left( 3 + \hbar \right)}(q^2)^2S_8
\end{eqnarray}
It should be noted that $S_5$ is a classical and quantum eigenfunction.
The twelfth eigenfunction $\phi_{12}$ is chosen to be a homogeneous one.

\subsubsection{$E_7$}
\label{calE7}
The generating function is defined in terms 
of the weights belonging to the {\bf 56}
dimensional representation:
\begin{equation}
G(x;\mathbf{56})=\prod_{\mu\in\mathbf{56}}(x+\mu\cdot q)
=\sum_{k=0}^{28} S_{2k} x^{56-2k}.
\label{E7calgen}
\end{equation}
 These are {\em minimal weights\/}
having the properties $\rho\cdot\mu=\pm1,0$ for $\forall\rho\in E_7$ and
\begin{equation}
\mu\cdot\nu=\left\{
\begin{array}{rl}
\pm3/2& \mu=\pm\nu,\\
\pm1/2 & \mbox{otherwise},
\end{array}
\right.\quad \mu, \nu\in\mathbf{56}.
\end{equation}
Moreover $\mathbf{56}$ is {\em even\/}, {\em i.e.\/} if $\mu\in\mathbf{56}$ then
$-\mu\in\mathbf{56}$. This is why the odd order terms in $x$ 
vanish in (\ref{E7calgen}). 
By applying the operator $\hat{A}_R$ (\ref{ARdef}) 
and using the above properties of
the weight vectors, we obtain
\begin{equation}
\hat{A}_RG={1\over2}\triangle G+{1\over4}\partial_x^2 G+{1\over{2x}}\partial_x G,
\label{e7calgeneq}
\end{equation}
or in terms of the $S_{2k}$,
\begin{equation}
\hat{A}_RS_{2k}={1\over2}\triangle S_{2k}+{1\over2}(29-k)(59-2k)S_{2k-2}.
\end{equation}
Some lower members of $S_{2k}(q_1,\ldots,q_7)$ are
\begin{equation}
S_0=1,\ S_2=-6q^2, \ S_4={33\over2}(q^2)^2,\ldots.
\end{equation}
The elementary excitations are for $N=$2, 6, 8, 10, 12, 14, 18, (see Table I) for which
$\{S_N\}$'s  are again functionally independent. 
The lower members of the corresponding
quantum eigenfunctions are:
\begin{eqnarray}
\phi_2&=&q^2-{7\over2}(18+\hbar),\\
\phi_6&=&S_6+{3(6970+609\hbar)/4}
\left[(q^2)^2-9(14+\hbar)/2(q^2-{7}(18+\hbar)/6)\right],\\
\phi_8&=&S_8+(555/2+165\hbar/4)S_6-315(1-\hbar)(q^2)^3\\
&&\quad + (536948+102416\hbar+4851\hbar^2)45/32 \left[
(q^2)^2-3(14+\hbar)(q^2-7(18+\hbar)/8)\right].\nonumber
\end{eqnarray}

\subsubsection{$E_8$}
\label{calE8}
We have not succeeded in deriving an equation for a generating function similar to
(\ref{acalgeneq}), (\ref{bcalgeneq}), (\ref{e6calgeneq}), (\ref{e7calgeneq}). 
We start from a Weyl invariant  power sum basis  in terms of roots
\begin{equation}
\Phi_{k}(q_1,\ldots, q_8)=\sum_{\rho\in\Delta_+}(\rho\cdot q)^{k},\quad k=2,8,12\ldots,
\end{equation}
for those eight $k$'s listed in the $E_8$ row of Table I. They are functionally
independent. For example,
$\Phi_2=30q^2$. Although the derivation of the classical and quantum eigenfunctions
for elementary excitations is straightforward, some results are too lengthy to present.
For want of proper and convenient notation, we show only some lower members of the
classical eigenfunctions:
\begin{eqnarray}
\varphi_2&=&q^2-120,\\
\varphi_8&=&\Phi_8 -197/750\, \Phi_2^3+7092/5\, \Phi_2^2-3404160\,\Phi_2+3063744000,\\
\varphi_{12}&=&\Phi_{12}-1473/20\,\Phi_8\Phi_2
+191/240000\,\Phi_2^5+132570\,\Phi_8+12551/5000\,\Phi_2^4
\nonumber\\
&&\qquad 118281/5\,\Phi_2^3+63871740\,\Phi_2^2-91975305600\,\Phi_2+551851833360000. 
\end{eqnarray}
Verification of the Propositions \ref{prop21}--\ref{prop23} in section two is easy but
tedious  calculation. 
\subsection{$F_4$}
\label{calF}
The long roots of $F_4$ are the roots of $D_4$
\begin{equation}
\Delta_L=\{\pm{\bf e}_j \pm{\bf e}_k,\
j,k=1,2,3,4|{\bf e}_j\in{\bf R}^{4}, {\bf e}_j\cdot
{\bf e}_k=\delta_{jk}\}
\end{equation}
and the short roots are the union of vector, spinor and anti-spinor weights of $D_4$:
\begin{equation}
\Delta_S=\{\pm{\bf e}_j |
j=1,2,3,4\}\cup\{(\pm{\bf e}_1\pm{\bf e}_2\pm{\bf e}_3\pm{\bf e}_4)/2 \}.
\end{equation}
We will consider a reduced theory in which $\omega=g_L=1$ and the short root coupling is
denoted by $g_S=\gamma$:
\begin{equation}
W=\sum_{\rho\in\Delta_{L+}}\log\rho\cdot
q+\gamma\sum_{\rho\in\Delta_{S+}}\log\rho\cdot q-{1\over2}q^2.
\end{equation}

Let us introduce the elementary symmetric polynomials in $\{q_1^2,q_2^2,q_3^2,q_4^2\}$, 
 as in the $B$ ($D$) series (\ref{genfun}):
\begin{eqnarray}
S_1&=&q_1^2+q_2^2+q_3^2+q_4^2\equiv q^2,\quad\quad 
S_2=q_1^2q_2^2+\ldots+q_3^2q_4^2,\nonumber\\
S_3&=&q_1^2q_2^2q_3^2+\ldots+q_2^2q_3^2q_4^2,\qquad\ \,
S_4=q_1^2q_2^2q_3^2q_4^2,
\end{eqnarray}
which are not Weyl invariant, except for $S_1$. A Weyl invariant basis for the elementary
excitations are for degree 2, 6, 8 and 12 (see Table I) polynomials \cite{turbf4}:
\begin{eqnarray}
\Phi_2&=&S_1, \quad \Phi_6=S_3-S_1 S_2/6,\quad \Phi_8=S_4-S_1 S_3/4+S_2^2/12,\nonumber\\
\Phi_{12}&=&S_4 S_2-S_2^3/36 -3 S_3^2/8 +S_1 S_2 S_3/8-3 S_1^2S_4/8.
\end{eqnarray}
The quantum eigenfunctions for the elementary excitations are:
\begin{eqnarray}
\phi_2&=& q^2-2\left(6(1+\gamma)+\hbar\right),\\
\phi_6&=&\Phi_6+\left(2(1+\gamma)+\hbar\right)/4\,\Phi_2^2\nonumber\\
&&\qquad\qquad-\left(2(2+\gamma)+\hbar\right)
\left(4(1+\gamma)+\hbar\right)/4\!\left[\vTm
3\Phi_2-2\left(6(1+\gamma)+\hbar\right)\right],
\\ 
\phi_8&=&\Phi_8+(3+\hbar)\,\Phi_6+(3+\hbar)
\left(2(2+\gamma)+\hbar\right)/8\,\Phi_2^2\nonumber\\
&&-
(3+\hbar)\left(2(2+\gamma)+\hbar\right)\left(4(1+\gamma)+\hbar\right)/8\left[\vTm 
2\Phi_2- \left(6(1+\gamma)+\hbar\right)\right],\\
\phi_{12}&=&\Phi_{12}+(3+2\hbar)\left(\vTs 2\Phi_8+(3+\hbar)\Phi_6\right)\!/2\,\Phi_2
-(3+2\hbar)\left(6(1+\gamma)+5\hbar\right)\Phi_8\nonumber\\
&&+(3+\hbar)(3+2\hbar)\left(2(2+\gamma)+\hbar\right)/24\Phi_2^3-
(3+\hbar)(3+2\hbar)\left(4(1+\gamma)+3\hbar\right)\Phi_6\nonumber\\
&&-(3+\hbar)(3+2\hbar)\left(2(1+\gamma)+\hbar\right)\left(2(2+\gamma)+\hbar\right)/16\!
\nonumber\\
&&\hspace*{15mm}\times
\left[\vTm 5\Phi_2^2-6\left(4(1+\gamma)+\hbar\right)\Phi_2
+ 2\left(4(1+\gamma)+\hbar\right)
\left(6(1+\gamma)+\hbar\right)\right].
\end{eqnarray}

\subsection{$G_2$ and Dihedral Root Systems}
\label{calDih}

The dihedral group of order $2m$, $I_{2}(m)$, is the group of orthogonal
transformations that preserve a regular $m$-sided polygon in two
dimensions. 
If all the roots are chosen to have the
same length $\alpha_{j}^{2}=1$, they can be parametrised as:
\begin{equation}
   \label{DihedralBasis}
   \alpha_{j}=\left(\cos(j\pi/m),\sin(j\pi/m)\right),\quad j=1,\ldots,2m.
\end{equation}
For odd $m$  all of the roots are in the same orbit of the reflection
group but for even $m$  there are two orbits, one consisting of the
$\alpha_{j}$ with odd $j$  and the other with even $j$.  Thus
 the $I_{2}(m)$ Calogero system has one coupling constant $g$ 
for odd $m$ and two couplings
$g_o$ and $g_e$
for even $m$ on top of the frequency $\omega$ of the harmonic confining potential.
The complete set of quantum eigenfunctions ($\hbar=1$) are given for all  
rank two Calogero
systems in \cite{kps}, with \(A_2\cong I_2(3)\),
\(B_2\cong I_2(4)\) and \(G_2\cong I_2(6)\).
So we concentrate on the elementary excitations with explicit $\hbar$ dependence.
The Coxeter invariant polynomials exist at degree 2  and \(m\) \cite{kps}:
\begin{equation}
 \Phi_2(q_1,q_2)=q^2,\quad  \Phi_m(q_1,q_2)=\prod_{j=1}^m(v_j\cdot q),
\end{equation}
where \(v_j\)'s are 
\begin{equation}
  v_j=(\cos((2j-1)\pi/2m),\sin((2j-1)\pi/2m)),\quad
   j=1,\ldots, m.
   \label{vmpara}
\end{equation}
If we introduce the two-dimensional polar coordinates system%
\footnote{We believe no confusion arises here, between the radial coordinate
variable \(r\) and the rank of the root system \(r\), which in this case is 2
of \(I_2(m)\).}
for \(q\),
\(
   q=r(\sin\theta,\cos\theta),
\)
the two Coxeter invariant polynomial variables read
\begin{equation}
   \Phi_2= q^2=r^2,\quad \Phi_m=2({r\over2})^m\cos m\theta.
\end{equation}
The essential part of the quantum eigenfunctions are a Laguerre polynomial
in $r^2$, (\ref{laguerrepoly}) times a Jacobi polynomial in $z=\cos m\theta$,  thus the
separation of variables is achieved.

As above, let us consider the reduced theory, $\omega=g=1$ for odd $m$ and $\omega=g_o=1$
and $g_e=\gamma$ for even $m$.
We have
\begin{equation}
\phi_2=\left\{
\begin{array}{lr}
\Phi_2-(m+\hbar),&\mbox{odd}\quad m,\\
\Phi_2-\left(\vTs m(1+\gamma)/2+\hbar\right),&\mbox{even}\quad m,
\end{array}
\right.\qquad 
\triangle(\Phi_2)^l=4l^2(\Phi_2)^{l-1},
\end{equation}
and
\begin{equation}
\hat{A}_R\Phi_m=\left\{
\begin{array}{lr}
0,&\mbox{odd}\quad m,\\
(\gamma-1)(\Phi_2)^{m/2-1}m^22^{-m},&\mbox{even}\quad m,
\end{array}
\right.\qquad 
\triangle\Phi_m=0.
\end{equation}
Thus $\Phi_m$ is a classical and quantum eigenfunction for odd $m$. 
For even $m$ we have
\begin{equation}
\phi_m=\Phi_m+{(\gamma-1)\over{2^{m-1}(\gamma+1+\hbar)}}r^m,
\end{equation}
which can be expressed as the Jacobi polynomial of degree one \cite{kps}.
The classical equilibrium point is
\begin{equation}
(\bar{r}^2,\bar{\theta})=(m,{\pi\over{2m}}), \quad
(m(1+\gamma)/2,{2\over m}\arctan\sqrt{\gamma}),
\end{equation}
for odd and even $m$, respectively. Verification of Proposition \ref{prop22}
is straightforward.
\subsection{H-Series}
\label{calH}
The non-crystallographic Coxeter groups of \(H_{3}\) and  \(H_{4}\)  are the
symmetry groups of the icosahedron and four-dimensional 600-cell, respectively.
The former consists of 30 roots and the latter 120. 
Let us start from  Coxeter invariant  power sum bases  in terms of roots
\begin{equation}
\Phi_{k}=\sum_{\rho\in\Delta_+}(\rho\cdot q)^{k},\quad k=2,6,10 \ \mbox{for}\ H_3;
\quad  k=2,
12, 20, 30 \ \mbox{for}\ H_4.
\end{equation}
The quantum eigenfunctions of the elementary excitations in $H_3$ are:
\begin{eqnarray}
\phi_2&=&q^2-3(10+\hbar)/2,\\
\phi_6&=&\Phi_6-{15}(13+3\hbar)/8\left[4(q^2)^2-(6+\hbar)
\left(\vTs 6q^2-3(10+\hbar)\right)\right],\\
\phi_{10}&=&\Phi_{10}-{(215+126\hbar)\over12}\left[4q^2-15(2+\hbar)\right]\Phi_6
+{5\over6}(58+63\hbar)\,(q^2)^4+{75\over4}(109+45\hbar)\,(q^2)^3\nonumber\\
&&-{25\over32}(10+3\hbar)(770+454\hbar+63\hbar^2)
\left[\vT 8(q^2)^2-(6+\hbar)\left(\vTs 10q^2-3(10+\hbar)\right)\right].
\end{eqnarray}
Some of the quantum eigenfunctions of the elementary excitations in $H_4$ are:
\begin{eqnarray}
\phi_2&=&q^2-2(30+\hbar),\\
\phi_{12}&=&\Phi_{12}-{315\over216}(565+66\hbar)
\left[\vTm
(q^2)^5-15(10+\hbar)
\left(\vTmm (q^2)^4-{20/3}(12+\hbar)\times\right.\right.\hspace*{1cm}\nonumber\\
&& \quad\quad\quad \left.\left.\times\left(\vTs
(q^2)^3-3(15+\hbar)\left((q^2)^2-{6/5}(20+\hbar)
(q^2-(30+\hbar)/3\right)\right)\vTmm\right)\vTm\right].
\end{eqnarray}
The other two eigenfunctions are too lengthy to be reported.
\section{Classical \& Quantum Eigenfunctions of the Sutherland Systems}
\label{trigs}
\setcounter{equation}{0}

As shown in \S\ref{roots} the eigenstates of the Sutherland system are
specified by {\em dominant highest weight\/}
$\lambda_{\vec{n}}$. The basis of the classical and quantum eigenfunctions are
thus the sum of the exponentials of $2i\mu\cdot q$ taken for the entire Weyl orbit
of 
$\lambda_{\vec{n}}$, $W\cdot\lambda_{\vec{n}}$, which we denote as
\begin{equation}
\Psi_{1^{n_1}2^{n_2}\cdots r^{n_r}}=\sum_{\mu\in W\cdot\lambda_{\vec{n}}}
e^{2i\mu\cdot q},\qquad 
\lambda_{\vec{n}}=\sum_{j=1}^rn_j\lambda_j,\quad 
n_j\in{\mathbb Z}_+.
\end{equation}
As usual, if the multiplicity $n_j$ is vanishing $n_j=0$, it is not written.
For example, the basis corresponding to the fundamental weights $\lambda_1$,
\ldots, $\lambda_r$ are $\Psi_1$, $\Psi_2$,\ldots, $\Psi_r$:
\begin{equation}
\Psi_1=\sum_{\mu\in W\cdot\lambda_{1}}e^{2i\mu\cdot q},\ \ldots, \
\Psi_r=\sum_{\mu\in W\cdot\lambda_{r}}e^{2i\mu\cdot q}.
\label{Psidefs}
\end{equation}
The operator $\hat{A}$ is lower triangular and the Laplacian \cite{kps,Kiri} is
diagonal in this basis:
\begin{eqnarray}
\hat{A}\Psi_{\lambda}=4\varrho\cdot\lambda \Psi_\lambda
+\sum_{|\lambda'|<|\lambda|}c_{\lambda'}\Psi_{\lambda'},
\qquad \triangle\Psi_\lambda=-4\lambda^2\Psi_\lambda.
\end{eqnarray}

As shown below there is a marked difference in the forms of the eigenfunctions
between the classical ($A$,
$B$, $C$ and $D$) and the exceptional ($E$, $F_4$ and $G_2$) root systems. 
The elementary excitations of the $A$-series Sutherland system can have the same 
classical and quantum eigenfunctions as in the Calogero case.

\subsection{A-Series}
\label{sutA}

The classical equilibrium of the $A_r$ Sutherland system is rather trivial
\cite{calpere,ahmcal}.
The prepotential  for the full and {\em reduced\/} theory read
\begin{equation}
W=g\sum_{j<k}^{r+1}\log\sin(q_j-q_k),\qquad
W=\sum_{j<k}^{r+1}\log\sin(q_j-q_k).
\end{equation}
The equations (\ref{Vmin}) determining the maximum of the ground state
wavefunction $\psi_0$ read
\[
\sum_{k\neq j}^{r+1}\cot[\bar{q}_j-\bar{q}_k]=0, 
\quad j=1,\ldots,r+1,
\]
which are satisfied by the {\em equally-spaced\/} configuration
$\bar{q}_j=(2j-(r+2))\pi/2(r+1)$. 
The Hessian $-\widetilde{W}$ has eigenvalues
$2\left\{ r, (r-1)2,
\ldots,(r+1-k)k, \ldots, 2(r-1), r\right\}$, which can be expressed as
$\{4\delta\cdot\lambda_1,\ldots,4\delta\cdot\lambda_k,\ldots,
4\delta\cdot\lambda_r\}$ with $\delta$ defined in (\ref{weylvec}). In these
formulas the trivial eigenvalue 0, coming from the  translational invariance, is
removed. The $k$-th eigenvector of 
$\widetilde{W}$ is simply
\(
v_k=(e^{2ik\bar{q}_1},\ldots, e^{2ik\bar{q}_{r+1}})
\).
The orthogonality condition of the eigenvectors $\{v_k\}$ read simply as
\(v_j\cdot v_k=\sum_{l}e^{2i(j+k)\bar{q}_l}=0\).

\bigskip
Let us introduce a generating function
\begin{equation}
G(x;\{e^{2iq_j}\})=\prod_{j=1}^{r+1}(x+e^{2iq_j})=
\sum_{k=0}^{r+1}S_k(\{e^{2iq_j}\})\,x^{r+1-k}.
\end{equation}
It is easy to see that the symmetric polynomial $S_k$ is equal to the basis 
$\Psi_k$ (\ref{Psidefs}) up to a term proportional to the ``center of mass"
$q_1+\cdots+q_{r+1}$ which is orthogonal to all the $A_r$ roots.
The generating function $G$ satisfies
\begin{equation}
\hat{A}G=2rx\partial_x G-2x^2\partial_x^2 G,\quad
\triangle G=-4(r+1)G+4x\partial_x G,
\label{sutAgen}
\end{equation}
which translate into
\begin{equation}
\hat{A}S_k=2k(r+1-k)S_k,\quad \triangle S_k=-4kS_k.
\end{equation}
Therefore
\begin{equation}
\varphi_k=S_k(\{e^{2iq_j}\}),\quad k=1,\ldots,r
\end{equation}
is a classical and quantum eigenfunction of the $k$-th elementary excitation with
eigenvalues $2k(r+1-k)$ and $2k\hbar(r+1-k+\hbar)$ of $\hat{A}$ (\ref{cleigen}) and
$\hat{H}$ (\ref{htilform}), respectively.
The absence of quantum corrections is a general property shared by eigenfunctions
belonging to {\em minimal weights\/} \cite{kps}. All the fundamental representations
of the A-series root systems are minimal.

\subsection{B- and C-Series}
\label{sutB}
Since the $B$ and $C$ root systems are closely related, $B\leftrightarrow C$
for $\alpha\leftrightarrow \alpha^\vee=2\alpha/\alpha^2$, many formulas for the
eigenfunctions etc take similar forms.
It is advantageous to write these expressions in parallel so that the
similarity and differences can be well appreciated.
The prepotentials for the  {\em reduced\/} theory read
\begin{eqnarray}
W&=&\sum_{j<k}^{r}\log(\cos2q_j-\cos2q_k)+\gamma\sum_{j=1}^{r}\log
\sin q_j,\qquad\ \, \mbox{B-series},\label{rWBsut}\\
&=&\sum_{j<k}^{r}\log(\cos2q_j-\cos2q_k)+\gamma\sum_{j=1}^{r}\log
\sin 2q_j,\qquad \mbox{C-series}.
\label{rWCsut}
\end{eqnarray}
The equations (\ref{Vmin}) determining the maximum of the ground state
wavefunction $\psi_0$ read
\begin{eqnarray}
\sum_{k\ne j}^r{1\over{\bar{x}_j-\bar{x}_k}}+
{\gamma\over{2}}{1\over{\bar{x}_j-1}}&=&0, \hspace*{34mm} \mbox{(B)},\\[-8pt]
&&\hspace*{24mm}j=1,\ldots,r,\nonumber\\[-8pt]
\sum_{k\ne j}^r{1\over{\bar{x}_j-\bar{x}_k}}+
{\gamma\over{2}}{1\over{\bar{x}_j-1}}+{\gamma\over{2}}{1\over{\bar{x}_j+1}}&=&0,
\hspace*{34mm} \mbox{(C)},
\end{eqnarray}
for $\bar{x}_j=\cos2\bar{q}_j$.
They determine $\{\bar{x}_j\}$ as the zeros of Jacobi polynomials \cite{cs}: 
\[
P_r^{(\gamma-1,-1)}(\bar{x}_j)=0 \quad  \mbox{(B)},\qquad
P_r^{(\gamma-1,\gamma-1)}(\bar{x}_j)=0 \quad 
\mbox{(C)}.
\]
Because of the identity
\[
P_r^{(a,-1)}(x)={a+n\over{2n}}(x+1)P_{r-1}^{(a,1)}(x),
\]
the (B) case always has one zero at $\bar{x}=-1$.
Let us choose $\bar{x}_r=-1\Leftrightarrow \bar{q}_r=\pi/2$.
The Hessian $-\widetilde{W}$ has eigenvalues
\begin{eqnarray}
2k(2r-k+\gamma-1),\quad &&k=1,\ldots,r-1,\quad \& \quad r(r+\gamma-1),\quad
\mbox{(B)},\\ 2k(2r-k+2\gamma-1),\quad &&k=1,\ldots,r,\hspace*{47mm}
\mbox{(C)},
\end{eqnarray}
in which the last one of the B-series belongs to the spinor representation
($\lambda_r$). The $k$-th eigenvector of $\widetilde{W}$ has a form
\begin{eqnarray}
\!\!v_k&=&\!\!\left(\sin2\bar{q}_1P_{k-1}(\bar{x}_1),
\ldots,\sin2\bar{q}_{r-1}P_{k-1}(\bar{x}_{r-1}),0\right),
\ k=1,\ldots,r-1,(r), \mbox{(B\&C)},
\label{sutBv1}\\
\!\!v_r&=&\!\!(0,0,\ldots,1), \hspace*{97mm} \mbox{(B)},
\label{sutBv2}
\end{eqnarray}
in which the polynomials $\{P_k\}$ of a single variable $x$ obey the three
term recursion relations:
\begin{eqnarray}
P_0(x)&=&1,\quad P_1(x)=x+{\gamma(\gamma-2)\over{(2r+\gamma-2)(2r+\gamma-4)}},
\label{sutBPrec1}\\
P_k(x)&=&\left(x+{\gamma(\gamma-2)
\over{(2r+\gamma-2k)(2r-2k+\gamma-2)}}\right)P_{k-1}(x)\nonumber\\
&&-\frac{4(r-k)(r-k+1)(r-k+\gamma)(r-k+\gamma-1)}{(2r-2k+\gamma)^2
(2r-2k+\gamma+1)(2r-2k+\gamma-1)}P_{k-2}(x),
\end{eqnarray}
for the B-series and
\begin{eqnarray}
P_0(x)&=&1,\quad P_1(x)=x,\\
P_k(x)&=&xP_{k-1}(x)-\frac{(r-k+1)(r-k+2\gamma-1)}
{(2r-2k+2\gamma+1)(2r-2k+2\gamma-1)}P_{k-2}(x),
\label{sutCPrec1}
\end{eqnarray}
for the C-series.
The orthogonality conditions for these discrete variable polynomials are
\begin{equation}
\sum_{j=1}^r(1-\bar{x}_j^2)P_k(\bar{x}_j)P_l(\bar{x}_j)=\delta_{k\,l},
\end{equation}
with $\{\bar{x}_j\}$ being the zeros of a Jacobi polynomial.

\bigskip
Let us introduce a generating function
\begin{equation}
G(x;\{\cos{2q_j}\})=\prod_{j=1}^{r}(x+\cos{2q_j})=
\sum_{k=0}^{r}S_k(\{\cos{2q_j}\})\,x^{r-k}.
\end{equation}
It is easy to see that the symmetric polynomial $S_k$ is proportional to the basis 
$\Psi_k$ (\ref{Psidefs}):
\begin{eqnarray}
S_k(\{\cos{2q_j}\})&=&2^{-k}\Psi_k,\quad k=1,\ldots,r-1, (r),\qquad\qquad
\mbox{(B\&C)},\\
\Psi_r&=&\sum_{\mu:spinor\ weights}e^{2i\mu\cdot q}=2^r\prod_{j=1}^r\cos q_j,
\hspace*{17mm} \mbox{(B)}.
\end{eqnarray}
The generating function satisfies
\begin{eqnarray}
\hat{A}G&=&2r(r+\gamma-1)G\,+\ 2(1-x^2)\partial_x^2G+2\gamma(1-x)\partial_xG,\quad\
\mbox{(B)},\\
\hat{A}G&=&2r(r+2\gamma-1)G+2(1-x^2)\partial_x^2G-4\gamma x\partial_xG,\qquad\quad\
\ \mbox{(C)},\\
\triangle G&=&-4rG+4x\partial_x G,\hspace*{59mm}\mbox{(B\&C)}.
\end{eqnarray}
These mean in turn
\begin{eqnarray}
\hat{A}S_k&=&
2k(2r-k+\gamma-1)S_k  
 +2\gamma(r-k+1)S_{k-1}\nonumber\\
&&\hspace*{40mm} +\ 2(r+1-k)(r+2-k)S_{k-2},  \quad\
\mbox{(B)},\\
\hat{A}\Psi_r&=&r(r+\gamma-1)\Psi_r,\hspace*{73.5mm}\mbox{(B)},\\
\hat{A}S_k&=&2k(2r-k+2\gamma-1)S_k+2(r+1-k)(r+2-k)S_{k-2},\quad\
\mbox{(C)},\\
\triangle S_k&=&-4kS_k,\quad \triangle\Psi_r=-r\Psi_r,\hspace*{52mm}\mbox{(B\&C)}.
\end{eqnarray}
The quantum eigenfunctions for the elementary excitations are:
\begin{eqnarray}
\phi_k&=&
S_k+\frac{\gamma(r-k+1)}{2r-2k+\hbar+\gamma}S_{k-1}\\
&&\quad+\frac{(r-k-1)(r-k+2)(2r-2k+\hbar+\gamma+\gamma^2)}
{2(2r-2k+\hbar+\gamma)(2r-2k+\hbar+\gamma+1)}S_{k-2}+\ldots,\quad k=1,\ldots,r-1,
\nonumber\\ 
\phi_r&=&\prod_{j=1}^r\cos q_j,\hspace*{103mm}\mbox{(B)},
\label{SutBspin}\\
\phi_k&=&\sum_{l=0}^{[k/2]} \frac{(r+2l-k)!\Gamma(r-k+\gamma+\hbar/2+1/2)}{4^l l!
(r-k)!
\Gamma(r-k+\gamma+\hbar/2+l+1/2)}S_{k-2l},
\quad k=1,\ldots,r,\quad \mbox{(C)}.
\end{eqnarray}
The corresponding eigenvalues are:
\begin{eqnarray}
2k\hbar(2r-k+\gamma-1+\hbar),\quad k&=&1,\ldots, r-1,\quad 
r \hbar(r+\gamma+\hbar/2-1),\quad \mbox{(B)},\\
2k\hbar(2r-k+2\gamma-1+\hbar),\quad k&=&1,\ldots, r,\hspace*{53mm} \mbox{(C)}.
\end{eqnarray}
The eigenfunction for the spinor weight in B-series 
$\phi_r$ (\ref{SutBspin}) has no 
quantum corrections.
The representations (\ref{sutBv1}), (\ref{sutBv2}) and the recursions
(\ref{sutBPrec1})--(\ref{sutCPrec1}) are obtained from these eigenfunctions.

\subsection{D-Series}
\label{sutD}
The reduced prepotential of the D-series Sutherland system is obtained by
removing the short (long) root coupling $\gamma$ terms from those of the B- and C-
series (\ref{rWBsut}), (\ref{rWCsut}).
This results in the change of the eigenvectors $\{v_k\}$ (\ref{sutBv1}),
(\ref{sutBv2}) $\to$ (\ref{sutDv1}),
(\ref{sutDv2}) and emergence of another eigenfunction associated with
anti-spinor weights $\phi_{r-1}$ (\ref{SutDspin}) which receives no quantum corrections.
The equations (\ref{Vmin}) determining the maximum of the ground state
wavefunction $\psi_0$ has a solution $\bar{q}_1=0$, $\bar{q}_r=\pi/2$ and
with $\bar{x}_j=\cos2\bar{q}_j$, $j=2,\ldots,r-1$ being the zeros of the Jacobi
polynomial $P^{(1,1)}_{r-2}(x)$ or equivalently of the Gegenbauer polynomial
$C^{3/2}_{r-2}(x)$:
\[
P^{(1,1)}_{r-2}(\bar{x}_j)=0,\quad C^{3/2}_{r-2}(\bar{x}_j)=0,\qquad j=2,\ldots,r-1.
\]
The Hessian $-\widetilde{W}$ has eigenvalues
\begin{eqnarray}
2k(2r-k-1),\quad k=1,\ldots,r-2,\quad \& \quad r(r-1)\ [2],
\end{eqnarray}
in which the exceptional one is doubly degenerate corresponding to the `fish tail'
of the $D$-series Dynkin diagram. 
The  corresponding eigenvectors of $\widetilde{W}$ are
\begin{eqnarray}
v_k&=&\left(0,\sin2\bar{q}_2P_{k-1}(\bar{x}_2),
\ldots,\sin2\bar{q}_{r-1}P_{k-1}(\bar{x}_{r-1}),0\right),
\quad k=1,\ldots,r-2,
\label{sutDv1}\\
v_{r-1}&=&(1,0,\ldots,0),\qquad v_r=(0,0,\ldots,1).
\label{sutDv2}
\end{eqnarray}
The polynomials $\{P_k\}$ of a single variable $x$ obey simple three
term recursion relations:
\begin{eqnarray}
P_0(x)&=&1,\quad P_1(x)=x,\\
P_k(x)&=&xP_{k-1}(x)-\frac{(r-k+1)(r-k-1)}
{4(r-k+1/2)(r-k-1/2)}P_{k-2}(x).
\end{eqnarray}

The generating function has the same form as in  the B, C cases:
\[
G(x;\{\cos{2q_j}\})=\prod_{j=1}^{r}(x+\cos{2q_j})=
\sum_{k=0}^{r}S_k(\{\cos{2q_j}\})\,x^{r-k}.
\]
It is easy to see that the symmetric polynomial $S_k$ is proportional to the basis 
$\Psi_k$ (\ref{Psidefs})  and that the two additional
bases are:
\begin{eqnarray}
S_k(\{\cos{2q_j}\})&=&2^{-k}\Psi_k,\quad k=1,\ldots,r-2, \\
\Psi_{r-1}&\propto&\prod_{j=1}^r\sin q_j,\quad
\Psi_{r}\propto\prod_{j=1}^r\cos q_j.
\label{spinant}
\end{eqnarray}
They satisfy
\begin{eqnarray}
\hat{A}G&=&2r(r-1)G\,+\ 2(1-x^2)\partial_x^2G,\\
\hat{A}\Psi_{r-1}&=&r(r-1)\Psi_{r-1},\qquad \hat{A}\Psi_{r}=r(r-1)\Psi_{r},\\
\triangle G&=&-4rG+4x\partial_x G,\quad \triangle\Psi_{r-1}=-r\Psi_{r-1},\quad
\triangle\Psi_{r}=-r\Psi_{r}.
\end{eqnarray}
These imply for $S_k$:
\begin{eqnarray}
\hat{A}S_k&=&2k(2r-k-1)S_k+2(r+1-k)(r+2-k)S_{k-2},\\
\triangle S_k&=&-4kS_k.
\end{eqnarray}
Thus we arrive at the quantum eigenfunctions corresponding to the elementary excitations:
\begin{eqnarray}
\phi_k&=&\sum_{l=0}^{[k/2]} \frac{(r+2l-k)!\Gamma(r-k+\hbar/2+1/2)}{4^l l!
(r-k)!
\Gamma(r-k+\hbar/2+l+1/2)}S_{k-2l},
\quad k=1,\ldots,r-2,\\
\phi_{r-1}&=&\prod_{j=1}^r\sin q_j,\qquad \phi_r=\prod_{j=1}^r\cos q_j,
\label{SutDspin}
\end{eqnarray}
with the eigenvalues of $\hat{H}$
\begin{equation}
2k\hbar(2r-k-1+\hbar),\quad k=1,\ldots, r-2,\quad 
r \hbar(r+\hbar/2-1)\ [2].
\end{equation}
Again, the eigenfunctions corresponding to the spinor and anti-spinor weights receive
no quantum corrections. These are minimal weights.
\subsection{E-Series}
\label{sutE}
For the Sutherland systems based on exceptional root systems, E, F and G,
the method of the generating functions seems not so useful 
as in the classical root systems cases, because of the `exceptional' character. 
The equilibrium points of the potentials are not related to known classical
polynomials in contrast to the cases discussed above.  
New polynomials describing the equilibria were introduced by Odake and Sasaki
\cite{cs}. Here we will construct the
eigenfunctions corresponding to the fundamental weights (elementary excitations)
starting from the basis $\Psi_1$, \ldots, $\Psi_r$ (\ref{Psidefs}). There is no
universally accepted way of naming the simple roots and fundamental weights of the
exceptional root systems. We show our conventions in terms of the Dynkin diagrams.

\subsubsection{$E_6$}
The symmetry of the Dynkin diagram is reflected in the structure of the 
eigenvalues and eigenfunctions, too. 
\begin{figure}
    \centering
\includegraphics{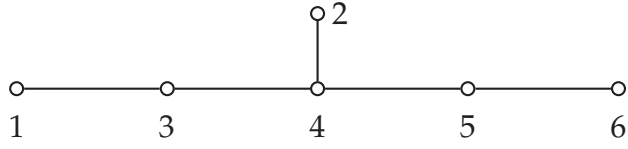}
   \caption{$E_6$ Dynkin diagram with the numbers of the simple roots attached.
}
    \label{fig:e6}
\end{figure}
The spectra of the Hessian $-\widetilde{W}$
and  the corresponding $\hat{H}$ 
in the order of $\lambda_1,\ldots,\lambda_6$ are:
\begin{eqnarray}
&&\{32,\, 44,\, 60,\, 84,\, 60,\, 32\},\\ 
&&\{{8\over3}\hbar(12+\hbar),\,
4\hbar(11+\hbar),\, {20\over3}\hbar(9+\hbar),\, 12\hbar(7+\hbar),\,
{20\over3}\hbar(9+\hbar),\, {8\over3}\hbar(12+\hbar)\}.
\end{eqnarray}
The  quantum eigenfunctions are listed in the order of increasing energy eigenvalues
and
the values of the
$\lambda_j^2$:
\begin{eqnarray}
4/3:&& \phi_1=\Psi_1,\hspace*{31mm} \phi_6=\Psi_6,
\label{e6mini}\\
2:&&\phi_2=\Psi_2+{72\over{(11+\hbar)}},
\label{e6roots}\\
10/3:&&\phi_3=\Psi_3+{40\over{(7+\hbar)}}\Psi_6,
\qquad\phi_5=\Psi_5+{40\over{(7+\hbar)}}\Psi_1,\\
6:&&\phi_4=\Psi_4+{24\over{(5+\hbar)}}\Psi_{16}+
{30(17+\hbar)\over{(5+\hbar)^2}}\Psi_2+
{720(17+\hbar)\over{(5+\hbar)^2(7+\hbar)}}.
\label{e64}
\end{eqnarray}
The orbits in sub-leading terms of an eigenfunction are contained 
in the {\em Lie algebra representation\/} 
specified by the dominant weight
of the leading term. Constant terms correspond to zero weights.
For example, the highest weight representation specified by $\lambda_4$
(\ref{e64}) consists of the Weyl orbits of $\lambda_4$, $\lambda_1+\lambda_6$,
$\lambda_2$ and some zero weights.
The lowest two eigenfunctions (\ref{e6mini}) consist of single orbits belonging to
the {\bf 27} and
$\overline{\bf 27}$ representations, which are minimal. 
Thus they do not receive quantum corrections.
They corresponds to the
left and right ends of the diagram, Fig.\ref{fig:e6}. 
The fundamental weight $\lambda_2$ (\ref{e6roots}) corresponds to the adjoint
representation, containing
all the roots and the rank number of zero weights. 
The constant term in (\ref{e6roots}), $72/(11+\hbar)$ reflects the number of roots
72 and the highest exponent 11 which is the `{\em height\/}' of the {\em highest
root\/},
that is $\lambda_2$ in the present case \cite{kps}. The longer the dominant weight
$\lambda_{\vec{n}}^2$ becomes, the more complicated structure has the
corresponding eigenfunction.
These are common features of all the eigenfunctions of the Sutherland
systems.

\subsubsection{$E_7$}
The spectra of the Hessian $-\widetilde{W}$ and the corresponding $\hat{H}$ are:
\begin{eqnarray}
&&\{68,\, 98,\, 132,\, 192,\, 150,\, 104,\, 54\},\\
&&\{4\hbar(17+\hbar),\, 7\hbar(14+\hbar),\, 
12\hbar(11+\hbar),\, 24\hbar(8+\hbar),\,15\hbar(10+\hbar)\nonumber\\
&& \hspace*{71mm}
8\hbar(13+\hbar),\, 3\hbar(18+\hbar)\},
\end{eqnarray}
which have no degeneracy. The Dynkin diagram has no symmetry.
\begin{figure}
    \centering
\includegraphics{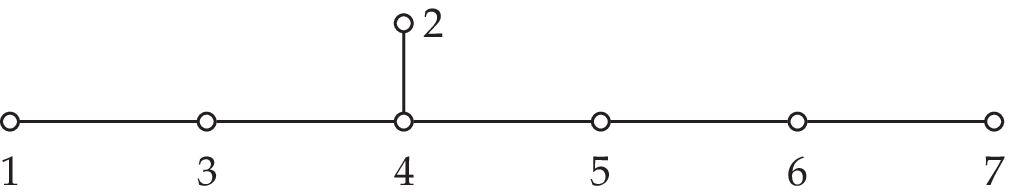}
   \caption{$E_7$ Dynkin diagram with the numbers of the simple roots attached.
}
    \label{fig:e7}
\end{figure}
The corresponding quantum eigenfunctions and the values of the $\lambda_j^2$ are:
\begin{eqnarray}
3/2:&&\phi_7=\Psi_7,\\
2:&&\phi_1=\Psi_1+{126\over{(17+\hbar)}},\\
7/2:&&\phi_2=\Psi_2+{72\over{(11+\hbar)}}\Psi_7,\\
4:&&\phi_6=\Psi_6+{60\over{(9+\hbar)}}\Psi_1+
{3780\over{(9+\hbar)(13+\hbar)}},\\
6:&&\phi_3=\Psi_3+{40\over{(7+\hbar)}}\Psi_6 +
{48(32+\hbar)\over{(7+\hbar)(8+\hbar)}}\Psi_1+
{2016(32+\hbar)\over{(7+\hbar)(8+\hbar)(11+\hbar)}},\\
15/2:&&\phi_5=\Psi_5+{40\over{(7+\hbar)}}\Psi_{17}+
{105(23+\hbar)\over{(7+\hbar)(13+2\hbar)}}\Psi_2
+{360(200+13\hbar)\over{(7+\hbar)(8+\hbar)(13+2\hbar)}}\Psi_7,\\
12:&&\phi_4=\Psi_4+{24\over{(5+\hbar)}}\Psi_{16}+
{30(17+\hbar)\over{(5+\hbar)^2}}\Psi_{27}+
{720\over{(5+\hbar)(7+\hbar)}}\Psi_{1^2}+
{720(17+\hbar)\over{(5+\hbar)^2(7+\hbar)}}\Psi_{7^2}\nonumber\\
&&+
{20(3340+911\hbar+68\hbar^2+\hbar^3)\over{(5+\hbar)^3(7+\hbar)}}\Psi_{3}+
{40(59325+19900\hbar+2126\hbar^2+80\hbar^3+\hbar^4)
\over{(5+\hbar)^3(7+\hbar)(11+2\hbar)}}\Psi_6\nonumber\\
&& +
{480(14735+3289\hbar+223\hbar^2+5\hbar^3)
\over{(5+\hbar)^3(7+\hbar)(11+2\hbar)}}\Psi_1+
{10080(1945+228\hbar+11\hbar^2)\over
{(5+\hbar)^3(7+\hbar)(11+2\hbar)}}.
\label{e74}
\end{eqnarray}
The first corresponds to the {\bf 56} dimensional representation
which is minimal. It has no $\hbar$ dependence. The second corresponds to the
set of roots (adjoint representation) with 126 roots and the highest exponent being 17.
The last expression (\ref{e74}) is much longer than its classical counterpart
\begin{eqnarray}
\varphi_4&=&\Psi_4+{24\over5}\Psi_{16}+{102\over5}\Psi_{27}+
{144\over7}\Psi_{1^2}+{2448\over{35}}\Psi_{7^2}+
{2672\over{35}}\Psi_3\nonumber\\
&&+{2712\over{11}}\Psi_6+{40416\over{55}}\Psi_1+
{112032\over{55}}.
\end{eqnarray}

For the $E_8$ eigenfunctions, we will present the classical ones 
simply because of the lack of space.

\subsubsection{$E_8$}
The spectra of the Hessian $-\widetilde{W}$ and the corresponding $\hat{H}$ 
in the order of $\lambda_1,\ldots,\lambda_8$ are:
\begin{eqnarray}
&&\{184,  \,272, \,364,\, 540,\, 440, \,336, \, 228,\,116\},\\
&&\{8\hbar(23+\hbar),\,  
16\hbar(17+\hbar),\,28\hbar(13+\hbar),\,60\hbar(9+\hbar) \nonumber\\
&&\hspace*{41mm}  40\hbar(11+\hbar),\,
24\hbar(14+\hbar),\,12\hbar(19+\hbar),\,4\hbar(29+\hbar)\},
\end{eqnarray}
which has no degeneracy. The Dynkin diagram has no symmetry.
\begin{figure}
    \centering
\includegraphics{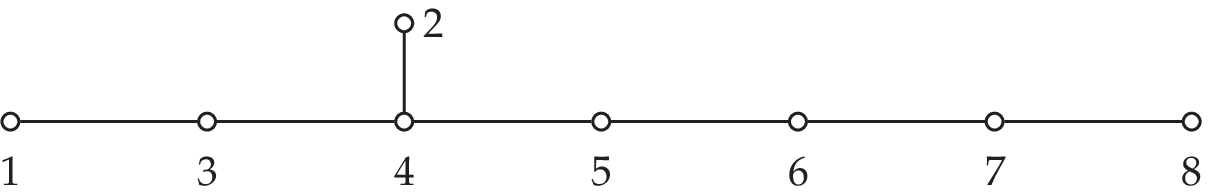}
   \caption{$E_8$ Dynkin diagram with the numbers of the simple roots attached.
}
    \label{fig:e8}
\end{figure}

The classical eigenfunctions for six lower elementary excitations are:
\begin{eqnarray}
2:&&\varphi_8=\Psi_8+{240\over{29}},\\
4:&&\varphi_1=\Psi_1+{126\over{17}}\Psi_8+{15120\over{17\cdot23}},\\
6:&&\varphi_7=\Psi_7+{84\over{11}}\Psi_1+{444\over{11}}\Psi_8+{35520\over{11\cdot19}},\\
8:&&\varphi_2=\Psi_2+{72\over{11}}\Psi_7+{4080\over 11^2}\Psi_1
+{215712\over{11^2\cdot13}}\Psi_8+{12942720\over{11^2\cdot13\cdot17}},\\
12:&&\varphi_6=\Psi_6+{20\over3}\Psi_{18}+{420\over 13}\Psi_{8^2}+
{203\over6}\Psi_2+{1776\over
13}\Psi_7+{361004\over{3\cdot13\cdot19}}\Psi_1\nonumber\\
&&\qquad\quad
+{4255608\over{11\cdot13\cdot19}}\Psi_8+{12660480\over{11\cdot13\cdot19}},
\\
14:&&\varphi_3=\Psi_3+{40\over7}\Psi_6+{192\over7}\Psi_{18}
+{1152\over 11}\Psi_{8^2}+{2608\over
23}\Psi_2+{12023496\over{7\cdot11\cdot17\cdot23}}\Psi_7\nonumber\\
&&\qquad\quad+{5525664\over{11\cdot17\cdot23}}\Psi_1
+{16392384\over{11\cdot17\cdot23}}\Psi_8+{592911360\over{11\cdot13\cdot17\cdot
23}}.
\end{eqnarray}
Most of the denominators contain the exponents of $E_8$, 
$\{1,7,11,13,17,19,23,29\}$. This is a common feature shared by
all the root systems but seen most clearly in the exceptional root systems cases.
The two unlisted eigenfunctions $\varphi_5$ and $\varphi_4$ have 
simply too many terms
to be presented here. For $\varphi_5$, $\lambda_5^2=20$, the number of
elements in the Weyl orbit of $\lambda_5$ is 241920 and the highest weight
representation of $\lambda_5$ is 146325270 dimensional.
The eigenfunction $\varphi_5$ contains 14 terms corresponding to the dominant
characters in the Lie algebra representation of the highest weight  
$\lambda_5$, that is
$\lambda_5$,
$\lambda_1+\lambda_7$, $2\lambda_1$, $\lambda_2+\lambda_8$,  $\lambda_7+\lambda_8$,
$\lambda_3$, $\lambda_6$, $\lambda_1+\lambda_8$, $2\lambda_8$, $\lambda_2$,
$\lambda_7$, $\lambda_1$, $\lambda_8$ and zero weights.
For $\varphi_4$, $\lambda_4^2=30$, the number of
elements in the Weyl orbit of $\lambda_4$ is 483840 and the highest weight
representation of $\lambda_4$ is 6899079264 dimensional.
The eigenfunction $\varphi_4$ contains 24 terms and some of their coefficients
are ratios of enormously large integers.
\subsection{$F_4$ and $G_2$}
\label{sutF}
The Sutherland systems based on $F_4$ and $G_2$ are interesting because of the
interplay of the long and short root couplings. 
While we show the Dynkin diagram of $F_4$ to indicate our convention of
the simple roots naming, we simply agree that $\alpha_1$ is the short simple root
of $G_2$, thus $\alpha_2$ is the long simple root.
\begin{figure}
    \centering
\includegraphics{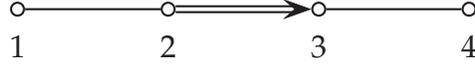}
   \caption{$F_4$ Dynkin diagram with the numbers of the simple roots attached.
}
    \label{fig:f4}
\end{figure}

The spectra of the Hessian $-\widetilde{W}$ and the corresponding $\hat{H}$ 
of $F_4$ in the order of $\lambda_1,\ldots,\lambda_4$ are:
\begin{eqnarray}
&&\{20 + 12\gamma, 
\,36 + 24\gamma, \,24 + 18\gamma, \,12 + 10\gamma\},\\
&&\{4\hbar(5+3\gamma+\hbar),\,12\hbar(3+2\gamma+\hbar),\, 6\hbar(4+3\gamma+\hbar),
2\hbar(6+5\gamma+\hbar)\}. 
\end{eqnarray}
The  quantum eigenfunctions are listed in the order of increasing energy eigenvalues
and
the values of the
$\lambda_j^2$:
\begin{eqnarray}
1:&&\phi_4=\Psi_4+{24\gamma\over{6+5\gamma+\hbar}},\\
2:&&\phi_1=\Psi_1+{6\gamma\over{4+\gamma+\hbar}}\Psi_4+
{24(4+\gamma+3\gamma^2+\hbar)\over{(4+\gamma+\hbar)(5+3\gamma+\hbar)}},\\
3:&&\phi_3=\Psi_3+{12\gamma\over{2+3\gamma+\hbar}}\Psi_1+
{12(2+5\gamma+6\gamma^2+\hbar+\hbar\gamma)\over{(3+2\gamma+\hbar)(2+\gamma+\hbar)}}
\Psi_4\nonumber\\
&&\hspace*{29mm} +{96\gamma(8+9\gamma+6\gamma^2+3\hbar+\hbar\gamma)\over{
(3+2\gamma+\hbar)(2+3\gamma+\hbar)(4+3\gamma+\hbar)}},\\
6:&&\varphi_2=\Psi_2+{4\gamma\over{2+\gamma}}\Psi_{14}
+{12(2+\gamma+\gamma^2)\over{
(2+\gamma)(3+\gamma)}}\Psi_{4^2} +{4\gamma(11+9\gamma)
\over{(2+\gamma)(3+\gamma)}}\Psi_{3}
\nonumber\\
&&\qquad +12\left[\vTs24+22\gamma+47\gamma^2+23\gamma^3\right]
\left[\vTs (2+\gamma)(3+\gamma)(4+3\gamma)\right]^{-1}\Psi_1
\nonumber\\
&&\
\qquad\qquad \ +24\gamma\left[\vTs 28+37\gamma+27\gamma^2\right]
\left[\vTs (2+\gamma)(3+\gamma)(4+3\gamma)\right]^{-1}\Psi_4
\nonumber\\
&&\ +96\left[\vTs
24+30\gamma+85\gamma^2+67\gamma^3+30\gamma^4\right]\!\!
\left[\vTs (2+\gamma)(3+\gamma)(4+3\gamma)(3+2\gamma)\right]^{-1}.
\label{f42}
\end{eqnarray}
Here we listed the classical eigenfunction $\varphi_2$ (\ref{f42}) for the highest
elementary excitation, simply for display reasons.

The spectra of the Hessian $-\widetilde{W}$ and the corresponding $\hat{H}$ 
of $G_2$  in the order of $\lambda_1$, $\lambda_2$ are:
\begin{eqnarray}
\{4 + {8\over3}\gamma, 
\,8 + 4\gamma\},\qquad
\{{4\over3}\hbar(3+2\gamma+\hbar),\,4\hbar(2+\gamma+\hbar)\}. 
\end{eqnarray}
The  quantum eigenfunctions are listed in the order of increasing energy eigenvalues
and
the values of the
$\lambda_j^2$:
\begin{eqnarray}
2/3:&&\phi_1=\Psi_1+{6\gamma\over{3+2\gamma+\hbar}},\\
2:&&\phi_2=\Psi_2+{6\gamma\over{3+\gamma+2\hbar}}\Psi_1
+{6(3+\gamma+2\gamma^2+2\hbar)\over
{(2+\gamma+\hbar)(3+\gamma+2\hbar)}},
\end{eqnarray}

\section{Summary and Comments}
\label{comments}
\setcounter{equation}{0}
The general theorem relating classical and quantum mechanics (section \ref{qmechanics})
is applied to the Calogero and Sutherland systems,
typical integrable multi-particle dynamics 
associated with root systems and having long range interactions.
The classical and quantum eigenfunctions for the elementary excitations are
constructed explicitly (section \ref{rationals}, \ref{trigs}), and their relation
to the eigenmodes of small oscillations (of the corresponding classical system)
is worked out in full. In particular, we obtain new representations for the
eigenvectors (of small oscillations) in terms of orthogonal polynomials of a
discrete variable (the discrete variable being the zeros of well-known classical
polynomials). It turns out that the quantum eigenfunctions are very closely
related to the classical counterparts. As a special case, the quantum
eigenfunction of the Sutherland system  belonging to a minimal representation
consists of a single Weyl orbit and it has exactly the same form as the classical
one, that is the quantum corrections are absent. The next simplest case, those
belonging to the adjoint representations is fully described by the number of
roots and the highest exponents. As shown in many explicit examples, the
classical and quantum eigenfunctions are fully described in terms of  the roots,
weights, exponents and characters, etc. We do believe that this is the case for
any eigenfunctions of the Calogero and Sutherland systems. To demonstrate this
assertion for any particular theory and   universally for all the Calogero and
Sutherland systems is a good challenge.

\section*{Acknowledgements}
I.L. is a post-doctoral fellow with the F.W.O.-Vlaanderen (Belgium).


\end{document}